\documentclass{aa} 

\usepackage{graphicx}
\usepackage{txfonts}
\usepackage{url}

\usepackage{subfig}

\usepackage{natbib,twoopt}
\usepackage[breaklinks=true]{hyperref} 
\usepackage{pdflscape}

\begin{document}

 \title{Taxonomic classification of asteroids based on MOVIS near-infrared colors}


 \author{M. Popescu \inst{1,2,3}
 \and
 J. Licandro \inst{1,2}
 \and
 J. M. Carvano \inst{4}
 \and
 R. Stoicescu \inst{3}
 \and
 J. de Le\'on\inst{1,2}
 \and
 D. Morate \inst{1,2}
 \and
 I.L. Boac\u{a} \inst{3}
 \and
 C. P. Cristescu \inst{5} 
 }

 \institute{ Instituto de Astrof\'{\i}sica de Canarias (IAC), C/V\'{\i}a L\'{a}ctea s/n, 38205 La Laguna, Tenerife, Spain
 \and
 Departamento de Astrof\'{\i}sica, Universidad de La Laguna, 38206 La Laguna, Tenerife, Spain
 \and
 Astronomical Institute of the Romanian Academy, 5 Cu\c{t}itul de Argint, 040557 Bucharest, Romania
 \and
 Observat\'{o}rio Nacional, rua Gal. Jos\'{e} Cristino 77, São Crist\'{o}v\~{a}o, 20921-400, Rio de Janeiro, Brazil
 \and
 Department of Physics, University Politehnica of Bucharest, Romania 
 }


 
 \abstract
 {The MOVIS catalog contains the largest set of near-infrared (NIR) colors for solar system objects. These data were obtained from the observations performed by VISTA-VHS survey using the Y, J, H and Ks filters. The taxonomic classification of objects in this catalog  allows to obtain large scale distributions for the asteroidal population, to study faint objects, and to select targets for detailed spectral investigations.
 }
 { We aim to provide a taxonomic classification for asteroids observed by VISTA-VHS survey. We derive a method for assigning a compositional type to an object based on its (Y-J), (J-Ks), and (H-Ks) colors.
 }
 {We present a taxonomic classification for 18\,265 asteroids from the MOVIS catalog, using a probabilistic method and the k-nearest neighbors algorithm. Because our taxonomy is based only on NIR colors, several classes from Bus-DeMeo were clustered into groups and a slightly different notation was used (i.e. the superscript indicates that the classification was obtained based on the NIR colors and the subscript indicates possible miss-identifications with other types). Our results are compared with the information provided by the Sloan Digital Sky Survey (SDSS) and Wide-field Infrared Survey Explorer (WISE).
 }
 {The two algorithms used in this study give a taxonomic type for all objects having at least (Y-J) and (J-Ks) observed colors. A final classification is reported for a set of 6\,496 asteroids based on the criteria that  KNN and probabilistic algorithms gave the same result, (Y-J)$_{err}\leq$ 0.118 and (J-Ks)$_{err}\leq$0.136. This set includes  144 bodies classified as $B_k^{ni}$, 613 as $C^{ni}$, 197 as $C_{gx}^{ni}$, 91 as $X_t^{ni}$, 440 as $D_s^{ni}$, 665 as $K_l^{ni}$, 233 as $A_d^{ni}$, 3\,315 as $S^{ni}$, and 798 as $V^{ni}$. 
 We report the albedo distribution for each taxonomic group and we compute new median values for the main types. We found that V-type and A-type candidates have identical size frequency distributions, but the V-types are five times more common than the A-types. Several particular cases, such as the A-type asteroid (11616) 1996 BQ2 and the S-type (3675) Kematsch, both in the Cybele population, are discussed.
 }
 
 \keywords{minor planets; techniques: photometric, spectroscopic; methods: observations, statistical}

\maketitle

\section{Introduction}

Spectro-photometric data in the near-infrared region have been used for a long time to characterize the surface composition of asteroids \citep[e.g.][]{1975ApJ...197..527J, 1982AJ.....87..834V, 1988Icar...74..454H, 2000Icar..146..161S, 2016AJ....151...98M}. Comparative planetology - interpretation of reflectance spectra by direct comparison with laboratory data, was one of the first tools used to analyze  the telescopic observations \citep[e.g.][]{1975ApJ...197..527J}.

Early works \citep[e.g.][]{1982AJ.....87..834V} showed that carbonaceous and silicate-like compositions can be determined based on the J, H, and Ks observations. As more data became available, evidences for several compositional groups appeared. These were represented by different taxonomic classes which occupy well defined regions in the near-infrared color-color plots \citep{1988Icar...74..454H}. Thus, spectro-photometry in the near-infrared region proved to be a method for obtaining compositional information for faint targets which are not available for spectroscopic studies. It has been recognized as one of the best current techniques for surveying a large number of distant populations such as Centaurs and trans-Neptunian objects \citep[e.g.][]{2010A&A...510A..53P}. 

The goal of asteroid taxonomy (the term is formed of the ancient Greek words \emph{taxis}$\sim$arrangement, and \emph{nomia}$\sim$method) is to identify groups of objects that have similar compositions. The classification schema is defined by statistically processing  spectro-photometric, spectral, and polarimetric data of a significant number of objects.

The first defined taxons (taxon - a taxonomic category) of asteroids were denominated by letters according to the compositional interpretation inferred from the observations in the visible wavelengths: C - carbonaceous, asteroids with observed properties similar to carbonaceous chondrite meteorites, and S - stony, asteroids which showed the distinct signature (i.e. 1 $\mu$m band) of pyroxene or olivine mixtures \citep{1975Icar...25..104C}. They noticed that more than $90\%$ of their observed minor planets fell into these two broad groups. This way of denominating groups continued as new data were reported -E for compositions like the enstatite meteorites, M for metallic asteroids and R for the objects with the reddest colors and moderately high albedos \citep{1978Icar...35..313B}.

The increasing amount of spectral data over the visible and near-infrared wavelength region allowed to extend the number of taxons by taking into account subtle features. Two of the latest and widely used taxonomies are \citet[][]{2002Icar..158..146B}, who defined  26 classes using data over the 0.44 - 0.92 $\mu$m spectral interval, and \citet[][]{2009Icar..202..160D}, who defined 24 classes based on data over the 0.45 - 2.45 $\mu$m wavelength range. The main groups initially defined translated into broad complexes (S-complex, C-complex, and X-complex) followed by end-member types (A, D, K, L, O, Q, R, and V). \citet{2010A&A...510A..43C} noted the fact that spectra similar to the templates proposed by the taxonomic systems were systematically recovered by independent authors using diverse data sets and different observing techniques, providing confidence in these systems. 

However, the classification accuracy in a taxonomic scheme depends on the observational data available. If only a part of the data is measured (i.e. spectrum over a short spectral interval, or just spectro-photometric data), it is still possible to assign a type, but the level of confidence decreases. Different methods can be applied. These methods range from forging directly the observational data into the schema to different parameterizations and transformations (e.g. principal component analysis) of the observed values. The problem of classifying new data can be summarized as: a) defining the parameters to be used; b) mapping the regions corresponding to different groups; and c) selecting a definition for the distance in this space.
 
The sky surveys provide a large amount of data for solar system objects \citep[e.g.][]{2000Icar..146..161S,2001A&A...375..275B,2001AJ....122.2749I,2016Icar..268..340C}. But, except for the spectroscopic surveys of asteroids \citep[e.g][]{1995Icar..115....1X,2002Icar..158..106B, 2004Icar..172..179L} which observed several thousands of objects, large observational datasets are limited to broad-band photometric filters. In this case a classification using an extended schema such as \citet{2009Icar..202..160D} is not possible. In order to retrieve the compositional information, several versions of taxonomies corresponding to broad classes were developed \citep[e.g.][]{2000Icar..146..161S, 2008DPS....40.6003M, 2010A&A...510A..43C, 2013Icar..226..723D,2016DPS....4832526R} in close relation with well know taxonomies \citep{1984PhDT.........3T,2002Icar..158..146B,2009Icar..202..160D}. One method for linking the colors with the spectral behavior for the SDSS data (the survey that provided the largest amount of data for solar system objects in the visible region) was to define the locus of each class in the parameter space using observations of previously classified asteroids (typically from SMASS and S3OS2 survey), meteorite spectra and synthetic spectra \citep{2010A&A...510A..43C}.

Taxonomic classification provides rough information about the surface composition of an asteroid. Accurate mineralogical characterization requires high quality spectra. However, from an observational point of view, assigning an object to a class based on spectro-photometric data has several advantages compared to spectral studies: requires less observing time, the faint targets (i.e. small objects or those at large heliocentric distances) can be characterized, and big datasets are available from the sky surveys \citep{2014Natur.505..629D}.

The aim of our work is to classify the asteroids based on the near-infrared colors provided by the MOVIS-C catalog \citep{2016A&A...591A.115P} using a schema fully compatible with the taxonomy defined by \citet{2009Icar..202..160D}. This study also proposes methods for investigating  future survey data such as the spectro-photometric observations of the solar system objects which will be observed by the Euclid survey \citep{2011arXiv1110.3193L,2018A&A...609A.113C}.

The MOVIS-C catalog was built based on VISTA-VHS survey and provides the largest set of near-infrared data up to now. The observations included three programs: (1) the VHS-GPS (Galactic Plane Survey), which uses the J and Ks filters;  (2) the VHS-DAS (Dark Energy Survey), which uses the J, H, and Ks filters; (3) the VHS-ATLAS, which uses the Y, J, H, and Ks filters \citep{2004SPIE.5493..411I,2010ASPC..434...91L, 2012A&A...548A.119C, 2013Msngr.154...35M, 2015A&A...575A..25S}. The number of measured colors for each solar system object varies according to the observing strategy and to the limiting magnitude of each filter. Also depending on the survey plan, the observations that enter in the computation of a color are distanced by different time intervals. The colors that are less affected are (Y-J), (J-Ks) and (H-Ks) for which the time span between two observations is about 7 min \citep[see][for a complete description]{2016A&A...591A.115P,Morate2018}. These colors were selected for assigning a taxonomic class. 

This paper is organized as follows: section 2 presents the data preparation and the framework for the classification schema. The methods used for the classification are described in section 3. The results are summarized in section 4 and discussed with respect to WISE albedo and SDSS data. Several implications of our findings are shown in section 5. A review of future perspectives is made in section 6 and conclusions are presented in section 7.

\section{Data preparation}

\subsection{Connecting the spectra to the near-infrared colors}

The first step for inferring the compositional information from  MOVIS-C NIR colors is to compute their relation with the spectral data. It was shown in the previous work \citep{2016A&A...591A.115P} that clear patterns appear in color-color plots of minor planets observed with the Y, J, H, Ks VISTA filters. The corresponding clusters can be linked to the taxonomic types defined by \citet{2009Icar..202..160D}.

The reflectance spectra of asteroids are obtained by dividing the observed spectral data of a minor planet by a G2V solar analog spectrum measured in similar conditions. The resulting curve is normalized (typically to be unity at 1.25 $\mu$m) because the shape of the features provides information about the surface composition. This technique is used for the V-NIR spectra where the flux of the reflected light is much larger compared to the emitted thermal radiation \citep[e.g.][]{Popescu2012}. Thus, the transformation of the reflectance spectra into colors and vice versa requires that the spectro-photometric data of G2V stars are taken into account.

Synthetic colors can be computed from reflectance spectra in the following manner: a) computing the flux ($F_f$) corresponding to a filter $f$ by multiplying the spectrum of asteroid ($S_{aster}$) by the transmittance curve of the filter - $H_f(\lambda)$, and by the spectrum of a standard G2V star ($S_{Sun}$), and integrating the result with respect to wavelength ($\lambda$); b) obtaining the corresponding colors ($C_{f1-f2}$ - where $f1$ and $f2$ are two different filters) by subtracting the magnitudes determined from the computed fluxes (this step involves a calibration constant - $C_{calib}$ according to the magnitude system). These steps are described by the equations Eq.~\ref{Spex2flux}  and Eq.~\ref{flux2color}.

\begin{equation}
F_f= $$\int H_f(\lambda) \cdot S_{aster}(\lambda) \cdot S_{Sun}(\lambda) d\lambda$$, \,\,\,f $$\in$$ \{Y,J,H,Ks\}
\label{Spex2flux}
\end{equation}

\begin{equation}
C_{f1-f2} = -2.5(\textrm{log}_{10}F_{f1}-\textrm{log}_{10}F_{f2}) + C_{calib}
\label{flux2color}
\end{equation}

By considering the reflectance flux $R_{f} = \int H_{f}\cdot R_{aster}d\lambda$, the  following approximation can be made made - Eq.~\ref{spec2color}, where $C_{f1-f2}^{Sun}$ is the color of the Sun corresponding to the filters $f1$ and $f2$. This introduces an error smaller than $\sim 0.007$ magnitudes (statistically computed  by taking into account a large number of spectra). This error can be neglected when compared to typical errors from the MOVIS-C catalog.
\begin{equation}
C_{f1-f2} = -2.5[\textrm{log}_{10}R_{f1} - \textrm{log}_{10}R_{f2}] + C_{f1-f2}^{Sun}
\label{spec2color}
\end{equation}
The advantage of this approximation is to consider the colors of the Sun which can be measured from the G2V solar analogs in the same conditions as the colors obtained for the asteroid data.  

\citet{2012ApJ...761...16C} provide the colors of the solar analogs based on the Two Microns All Sky Survey (2MASS). We searched the VISTA observations for G2V stars identified based on 2MASS. After having removed the outliers (i.e. colors outside $3\sigma$ interval from the median values) we took the medians and found the following values: $(Y-J)_V = 0.219 $, $(J-H)_V = 0.262 $, (J-Ks)$_V = 0.340 $, $(H-Ks)_V = 0.079 $. The comparison with the values reported  by \citet{2012ApJ...761...16C} and transformed to VISTA filter system (see \cite{2016A&A...591A.115P} for the computation) shows the largest difference for the (Y-J) color (about 0.013). This can be attributed to the fact that the 2MASS survey did not use the Y filter and its value is extrapolated from the other three filters. Thus, the colors of G2V solar analogs used in this article are: $(Y-J)_V = 0.219 $ - resulting form current determination, $(J-H)_V = 0.255 $, (J-Ks)$_V = 0.336 $, $(H-Ks)_V = 0.082 $ - resulting from \citet{2012ApJ...761...16C} after applying the transformation to VISTA filters.

Equation~\ref{spec2color} can be reverted to obtain reflectances from colors:
\begin{equation}
{R_{aster}^{f1}}/{R_{aster}^{f2}} = 10^{-0.4(C_{f1-f2} - C_{f1-f2}^{Sun})}
\label{color2reflec}
\end{equation}
In this case, the normalization can be made relative to the J filter ($R_{aster}^{J} = 1$), which is consistent to spectral data. This relation is useful for describing the spectral parameters (e.g. slopes) based on the MOVIS-C colors and for comparing them with existing data in the literature.

\subsection{The reference colors}

The taxonomic system defined by \citet{2009Icar..202..160D} is one of the most used for classifying asteroids using their reflectance spectra in the visible and near-infrared wavelength regions. The authors define 24 classes which closely follow the visible wavelength taxonomy of \citet{2002Icar..158..146B} and the one of \citet{1984PhDT.........3T}. The taxons were obtained by applying Principal Component Analysis (PCA) to a set of 371 spectra of asteroids for which the slope had been removed. The boundaries between classes were determined in the PC1-PC5 space. The reflectance values of the standard spectral types are defined for 41 wavelengths, equally distanced at 0.05 $\mu$m, over the 0.45 to 2.45 $\mu$m interval.

In order to obtain a classification compatible with the one of \citet{2009Icar..202..160D} the same spectral set has to be used as reference. Thus, we applied the transformation shown by Eq.~\ref{spec2color} to the 371 spectra used to define the \citet{2009Icar..202..160D} taxonomic system. For each class we obtained the mean value and the standard deviation of (Y-J), (J-Ks), and (H-Ks) colors. These are summarized in Table~\ref{TaxonColor}. All standard deviations smaller than 0.02 were limited to this value to account for other type of errors (i.e. approximations made by color computation, light-curve variations, solar color errors, etc). The reflectance in the Y, J, H, and  Ks filters for each standard spectral curve are shown in Annex (Fig.~\ref{ReflectanceClasses}).

Note that some of the classes in the \citet{2009Icar..202..160D} taxonomy are defined based on a very low number of objects. This is the case for the A-types (6 objects), B-types (4 objects), Sa-types (2 objects), Sv-types (2 objects), and even Cg-, O-, and R-types defined with a single object. Due to this fact their limits are poorly constrained, with the standard deviations being too large (A-types), too small (Sa-, Sv-types), or undefined (for those classes represented by a single object). This uncertainty translates in the color domain, i.e. the color intervals covered by these types are insufficiently defined.

The ability of the VISTA filter set to distinguish between taxonomic classes can be assessed by looking at the distribution of the colors in each class. To illustrate the information contained by the (Y-J), (J-Ks), and (H-Ks) colors we plot the average values and the standard deviation obtained for each color and spectral class (Fig.~\ref{SingleColorDistrib}). These plots show that the end-members A, B, and V types can be inferred by a single color.  For example, the (Y-J) color allows the following assumptions: 1) because of the steepness of 1 $\mu$m band, an object with (Y-J) > 0.45 is an A-type, an O-type, an R type, or a V-type; 2) the V-types (within 2$\sigma^V_{(Y-J)}$, where $\sigma^V_{(Y-J)}$ is the standard deviation of V-type template) have (Y-J) $\geq~\sim$0.5; 3) in general, the S-complex has larger values of (Y-J) than the C-complex, with a relative border between them at $\sim$ 0.3; 4) the B-types have negative NIR slopes, and they occupy the region with (Y-J)$\leq$0.219, which is the (Y-J) color of the Sun. A value of (J-Ks) $> 0.8$ indicates an extremely red near-infrared spectrum corresponding to A- or D-types.

\begin{figure}
\begin{center}
\includegraphics[width=9cm]{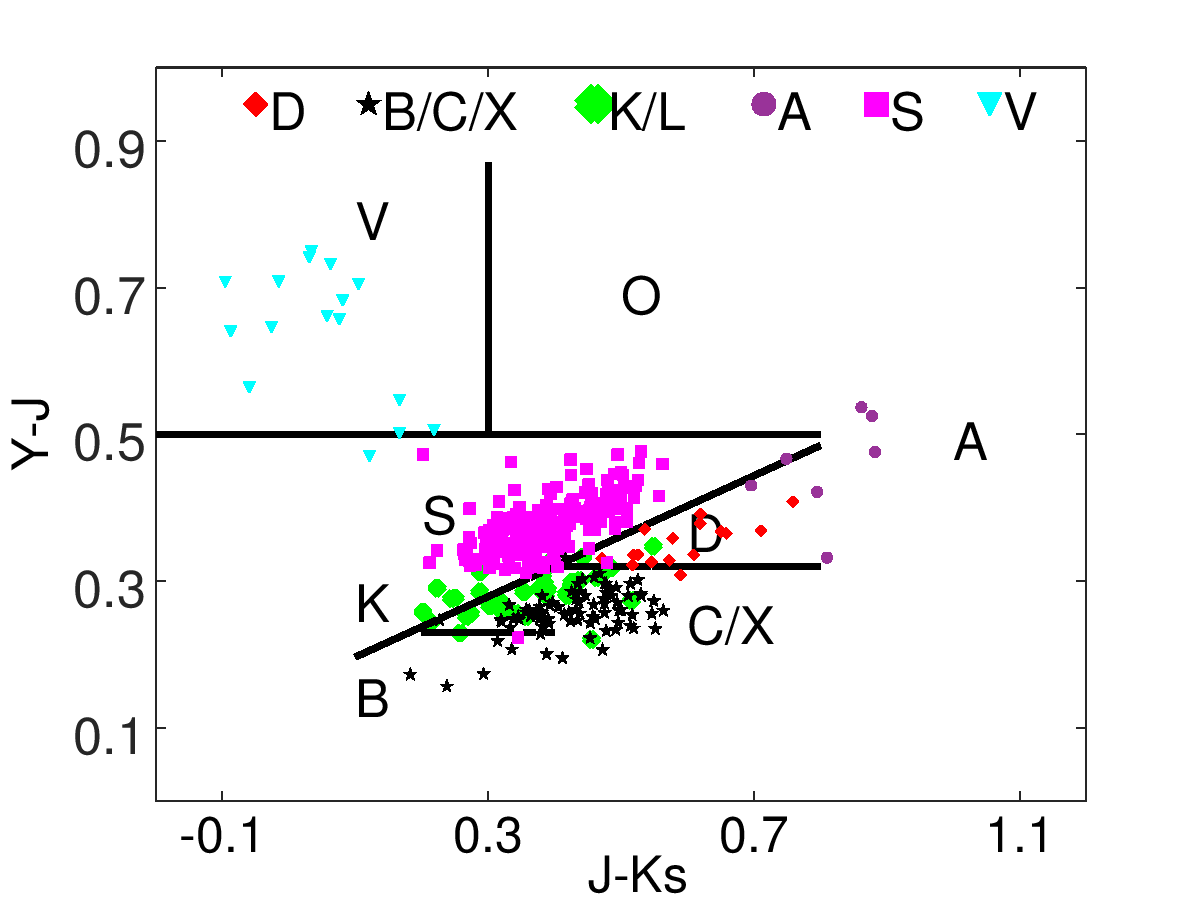}
\caption{(Y-J) vs. (J-Ks) color-color plot derived from the 371 spectra used to define the \citet{2009Icar..202..160D} taxonomy. The main taxonomic types are plotted with different colors for comparison. A-types include A- and Sa-types; C and X refer to C- and X-complexes; S-types include the S-complex and Q-types. We indicate the limits for each taxonomic class assumed in this work with straight lines.}
\label{CompColor371Spectra}
\end{center}
\end{figure}

The (Y-J) vs (J-Ks) contains most of the information for separating the main taxonomic groups \citep{2016A&A...591A.115P}. This result is emphasized by plotting the computed colors for the reference set (Fig.~\ref{CompColor371Spectra}). Several remarks can be outlined: 1) the region (Y-J)$\geq 0.5$ and (J-Ks)$\leq 0.3$ is distinctive \citep{2017A&A...600A.126L} and was identified as corresponding to the V-type asteroids; 2) the O and R types (their spectra show a deep 1 $\mu$m absorption band and a shallow 2 $\mu$m feature) are in the interval  (Y-J)$\geq 0.5$ and $0.7 \geq(J-Ks)$ $\geq 0.3$; 3) the S-complex can be separated from the C-complex and the D-type based on the line $\alpha_{(Y-J)} = 0.412^{\pm0.046}\cdot(J-Ks) + 0.155^{\pm0.016}$ \citep{2016A&A...591A.115P}. This line marks the separation between the groups associated to silicaceous and carbonaceous compositions, but it cuts the K-, L- type regions which span intermediate color values. Using the (Y-J) value, we can further divide the area below the $\alpha$ line as follows: D-types have (Y-J) $\geq0.3$, B types have (Y-J)$\leq0.219$ and (J-Ks)$\leq0.34$, while the C- and X- complexes are located between these two borders.

This description is an indication based on the computed colors of the reference spectral set.  The real data deals with photometric errors, phase angle effects or reddening due to space-weather, and light-curve variations which can change the categorization. To overcome this fact, we clustered the 24 classes defined by \citet{2009Icar..202..160D} into several groups by considering similar compositional types and their distance in the \{(Y-J), (J-Ks), (H-Ks)\} parameter space. These groups are considered for classifying the observational data provided by the MOVIS-C catalog. They are described below by showing the analogous Bus-DeMeo types. The superscript index of each letter indicates that the group was defined based on NIR colors ($ni$) and the subscript indicates some possible confusions (the level of accuracy will be discussed in the following sections). 

\begin{itemize}
 \item \emph{$B_k^{ni}$} = (B), this type is characterized by NIR color values lower than those of the Sun. It is distinctive from the C-complex objects. A result found after studying the classified sample was that the close borders with the K-types may give some confusion with this one (i.e. there are K-types with blue NIR colors).
 \item \emph{$C^{ni}$} = (C, Cb), has higher (J-Ks) and (H-Ks) values compared to the Cg,Cgh,Ch (Table~\ref{TaxonColor}). This allows us to split the C-complex into two groups. However, the distance between them in the color space is small (less than 0.1 mag.), and for the separation to be effective it requires accurate photometry. As it will be shown in Section 4 this group identifies with high probability ($\geq85\%$) to the low albedo asteroids.
 \item \emph{$C_{gx}^{ni}$} =(Cg,Cgh,Ch,Xc,Xe), the near-infrared colors are unable to separate some classes of the C-complex from those of the X-complex, so this group includes both low and high albedo asteroids. The colors corresponding to this group occupy an intermediate region between $C^{ni}$ and {$X_t^{ni}$}. 
 \item \emph{$X_t^{ni}$} = (T, X, Xk), corresponds to objects with the highest color values from the X-complex. It also includes both low and high albedo asteroids.
 \item \emph{$D_s^{ni}$} = (D) has the largest spectral slope in the NIR, compared to {$X_t^{ni}$}. The definition of \citet{2009Icar..202..160D} is based on the V-NIR spectral data of 16 objects and it covers broad reflectance values. The reddest objects belonging to this group can be confused with the A, Sa types and marginally with Sr.
 \item \emph{$K_l^{ni}$}, the K and L types have intermediate color values between C- and S- complexes in  the  NIR colors space. Based only on (Y-J), (J-Ks), and (H-Ks), the asteroids classified in this group can be misidentified with some of the S-types or C-types. The group also serves for reducing the misidentifications between the silicaceous and carbonaceous compositions associated to the C and S complexes.
 \item \emph{$A_d^{ni}$} = (A,Sa), it is characterized by the reddest (J-Ks) color, a moderate to high (Y-J) color, and a moderate (H-K). The borders of the group are insufficiently determined in the color space (e.g. large $\sigma_{(J-Ks)} = 0.152$ value) due to the low number of reference objects used for the definition of these classes (six for A and two for Sa ), and because they are widely spread in the color space. The asteroids assigned to this group based on MOVIS-C must have an extremely red NIR spectrum. A separation between $A_d^{ni}$ and D-types can be made by including the albedo value. This was unexpected from the investigation of the reference set which does not include D-types with so red values. 
 \item \emph{$S^{ni}$} = (Q,S,Sq,Sv,Sr), it corresponds to silicate-like compositions which show spectral bands at 1 $\mu$m and 2 $\mu$m. 
 \item \emph{$V^{ni}$} = (V), is distinctive in the (Y-J) versus (J-Ks) color-color space \citep{2017A&A...600A.126L}. It corresponds to compositions similar to howardite-eucrite-diogenite meteorites (HED).
\end{itemize}

We note that the O and R types were not considered, as they are defined based on a single object. It is shown in the discussion section that there are several tens of objects with similar color values as these types. The L type is included in the same group as K-type.

\begin{table*} 
\caption{Characteristics of the selected sample. The upper error limits for the quartiles are shown. The sample size $N_{YJKs}$  - the number of objects having at least the (Y-J) and (J-Ks) colors determined and for which both errors are in the given limits, and  $N_{YJHKs}$ - the number of objects with the (Y-J), (J-Ks) and (H-Ks) having all three errors in the limits, are provided. The percentages are shown relative to the total number of objects reported in the MOVIS-C catalog.}
\centering
\begin{tabular}{l |c c c | c c |c c}\hline\hline
 & $(Y-J)_{errlim}$  & (J-Ks)$_{errlim}$ & $(H-Ks)_{errlim}$ & $N_{YJKs}$ & $[\%]_{YJKs}$ & $N_{YJHKs}$ & $[\%]_{YJHKs}$ \\ \hline
Q1       & 0.059& 0.073& 0.079 & 4483     & 8            &  1866       & 3 \\
Q2       & 0.118& 0.136& 0.146 & 9097     & 17           &  3828       & 7 \\
Q3       & 0.199& 0.215& 0.230 & 13695    & 26           &  5881       & 11 \\
Q4       & 0.576& 0.562& 0.570 & 18265    & 34           &  8027       & 14 \\
\hline
\end{tabular}
\label{Quartile}     
\end{table*}

\subsection{MOVIS-C data selection and ranking}

The VHSv20161007 data release contains 141132 logs of stack-frames corresponding to the observations acquired between November 4, 2009 and March 27, 2016. We obtained an updated version of MOVIS-C catalog containing the near-infrared colors for 53\,447 solar system objects by running the pipeline described in \cite{2016A&A...591A.115P} for this set of VISTA-VHS observations. The search was performed considering the objects listed in the ASTORB version of 2017, March. We recovered color information for 57 NEAs (near-Earth asteroids), 431 Mars Crossers, 612 Hungaria asteroids, 51382 main-belt asteroids, 218 Cybele asteroids, 267 Hilda asteroids, 434 Trojans, and 29 Kuiper Belt objects.

The details regarding the survey and the pipeline used to retrieve the colors from the observations catalogued by VISTA Science Archive which is part of VISTA Data Flow System \citep{2004SPIE.5493..401E,2004SPIE.5493..423H,2004SPIE.5493..411I, 2010ASPC..434...91L} are described by \cite{2016A&A...591A.115P} and the references herein. Here we provide here a brief summary. The magnitudes and fluxes were retrieved from the \emph{ vhsDetection} database\footnote{\url{http://horus.roe.ac.uk/vsa/www/vsa_browser.html}}, which contains the sources from each individual stack frame. This type of image is obtained by adding typically two exposures of 7 or 15 seconds each (depending on the program). The derived colors are the differences of the average magnitudes (typically there were at least two detections for each band) in each filter, after the outliers and spurious data have been removed. 

The MOVIS-C colors catalog contains photometric observations with errors varying from accurate measurements (i.e photometric errors less than 0.01 magnitude) up to the detection limit of the pipeline (i.e. about $\approx$0.4 magnitude error). A first approach for accounting the uncertainties is to split the MOVIS-C catalog in four subsets using the quartiles defined on the distributions of the color magnitude errors for the (Y-J), (J-Ks) and (H-Ks).  The quartiles are given by the thresholds that divide the dataset into four equal-size groups with respect to error.  For example, the second quartile cut-off value corresponds to the median of errors. The sub-sets are denoted by Q1  - lower quartile, Q2 - median, and Q3 - upper quartile. Table~\ref{Quartile} gives the upper error limits and the number of objects. The subsets are defined considering objects with measurements in three or four filters and with errors for the colors lower than the thresholds of the quartiles. 

There are 37\,634 objects with determined (Y-J) color,  31\,750 with (J-Ks), and 11\,268  with (H-Ks). The (Y-J) and (J-Ks) color space contains most of the information with respect to taxonomic type. There are 18\,265 asteroids having both (Y-J) and (J-Ks) measured. In the best case when an object was observed with all four filters, we chose the (Y-J), (J-Ks), and (H-Ks) colors for the classification because they were obtained in similar time intervals of about $\sim$ 7 min \citep{2016A&A...591A.115P}. 

In order to evaluate the possible errors introduced by the lightcurve variation we use the data available in the Asteroid Lightcurve Database\footnote{\url{http://www.minorplanet.info/lightcurvedatabase.html}}. The median rotation period for more than 18\,000 asteroids is $\sim$6.3 h, and the median of the light curve amplitudes is $\sim$0.38 magnitudes. Statistically, these values translate into an uncertainty of $\sim$0.03 magnitudes for (Y-J) and (J-Ks) which have about a $\sim$7 min time interval between the observations with each filter.

\section{Methodology}

The classification of asteroids based on colors is sensitive to magnitude uncertainties. The fact that the spectral interval is sampled by few broad-band filters makes the uncertainties substantially account for the shape of the equivalent spectrum. In this context we used two approaches  for classifying MOVIS-C data: a probabilistic one following a similar schema  to \citet{2010A&A...510A..43C} -- which was  successfully applied for SDSS data,  and a second one based on machine-learning algorithms provided  with the \emph{scikit-learn}\footnote{\url{http://scikit-learn.org/}} module for Python \citep{scikitlearn,sklearnapi}. The results reported by using both methods are provided in the catalog accompanying the paper.

For implementing the probabilistic approach, we considered a normal distribution for each taxonomic type defined by \citet{2009Icar..202..160D} in the \{(Y-J), (J-Ks), (H-Ks)\} colors space. The distributions are defined based on the mean and standard deviation of colors (Table~\ref{TaxonColor}) computed from the 371 spectra used for the definition of Bus-DeMeo taxonomy. The computation is performed as explained in Section 2.2. The normal distribution is widely used in the description of the natural phenomena. In this case there is no physical evidence supporting a particular type of distribution. This choice is made in a convenient way as it does not require to define borders in the parameter space and outlines the continuity of the spectral features shown in the color-color plots. This continuity of spectral features was also observed with the visible colors provided by the SDSS data \citep{2015A&A...577A.147H}.

For each observed colors we attributed a normal distribution around the nominal value with a standard deviation equal to the uncertainty of the measurement. The probability for the asteroid color of falling into the limits of a given class is calculated as the area of the distribution that overlaps the class distribution (the class is modeled as a Gaussian distribution with mean and standard deviation shown in Table~\ref{TaxonColor}), and the probability of the observation belonging to a taxonomic type is the product of the probabilities of all colors. This method provides a measure of how much a set of observations resembles to a given template, which in this case is one of the 24 types defined by the Bus-DeMeo taxonomy. The class with the highest probability points to the taxonomic group assigned to the object.

A second approach for the classification schema is to use different machine-learning methods. A similar approach has been used by  \citet{2016AJ....151...98M} for classifying near-Earth objects observed with Z, J, H and K filters. Four our analysis, we used the computed colors of the spectra presented by \cite{2009Icar..202..160D} as the training set.  The  following methods were tested: Extreme Gradient Boosting (XGB), Random Forests 
(RF) with different parameters, and Nearest Neighbors (kNN). 

\begin{table} 
\caption{Comparison between different classification algorithms based on the "leave-one-out" test performed with the computed colors from the spectra used for the definition of Bus-DeMeo taxonomy. $Acc_{YJHKs} $ is the accuracy obtained when running the algorithm with all three colors, and $Acc_{YJKs}$ is the accuracy when only (Y-J) and (J-Ks) are used.}
\centering
\begin{tabular}{l c c}\hline\hline
Algorithm & $Acc_{YJHKs}[\%]$ & $Acc_{YJKs}[\%]$ \\ \hline
XGB       & 87.8              & 84.6 \\
RF leaf 1 & 86.2              & 85.4 \\
RF leaf 2 & 86.2              & 85.6 \\
RF leaf 3 & 85.6              & 85.4 \\
RF leaf 4 & 85.4              & 84.6 \\
kNN 3     & 87.0              & 84.6 \\
kNN 5     & 87.0              & 85.9 \\
\hline
Probabilistic      & 80.2              & 80.2\\
\hline
\end{tabular}
\label{Algorithms}     
\end{table}

To validate the algorithms we performed the "leave-one-out" cross-test with respect to the training set. The test involves using one object from the reference set for validation and the remaining data as the training set. This is repeated for all objects in the training set. Table ~\ref{Algorithms} compares the accuracy of classifications for these algorithms, considering two different cases: the first case assumes the three color observations \{(Y-J),(J-Ks),(H-Ks)\} and the second case assumes two color measurements \{(Y-J),(J-Ks)\}. The accuracy results are comparable in the limit of 2$\%$. Similar results were obtained by \cite{2016AJ....151...98M} using  the (Z - J), (J - H), (J - Ks) colors based on a training sample of 319 synthesized asteroid colors.

We report the result of classification using the kNN (k=3, 3NN) algorithm. Although 5NN performs slightly better than 3NN according to the "leave one out" test, due to the fact that for end member classes we have a very small number of training samples (A - 6 objects, B-4 objects, D-16 objects) we decide to use the last one. This simple and effective method works well for identifying peculiar types. To account for the magnitude errors we used a Monte-Carlo approach: each measurement was randomized $10^6$ times considering a Gaussian distribution with the color as the mean value and color error as the standard deviation. We counted the frequency of classification for each group to obtain the classification probabilities and we adopted the most likely taxonomic class for the target.

\begin{figure}
\begin{center}
\includegraphics[width=9.5cm]{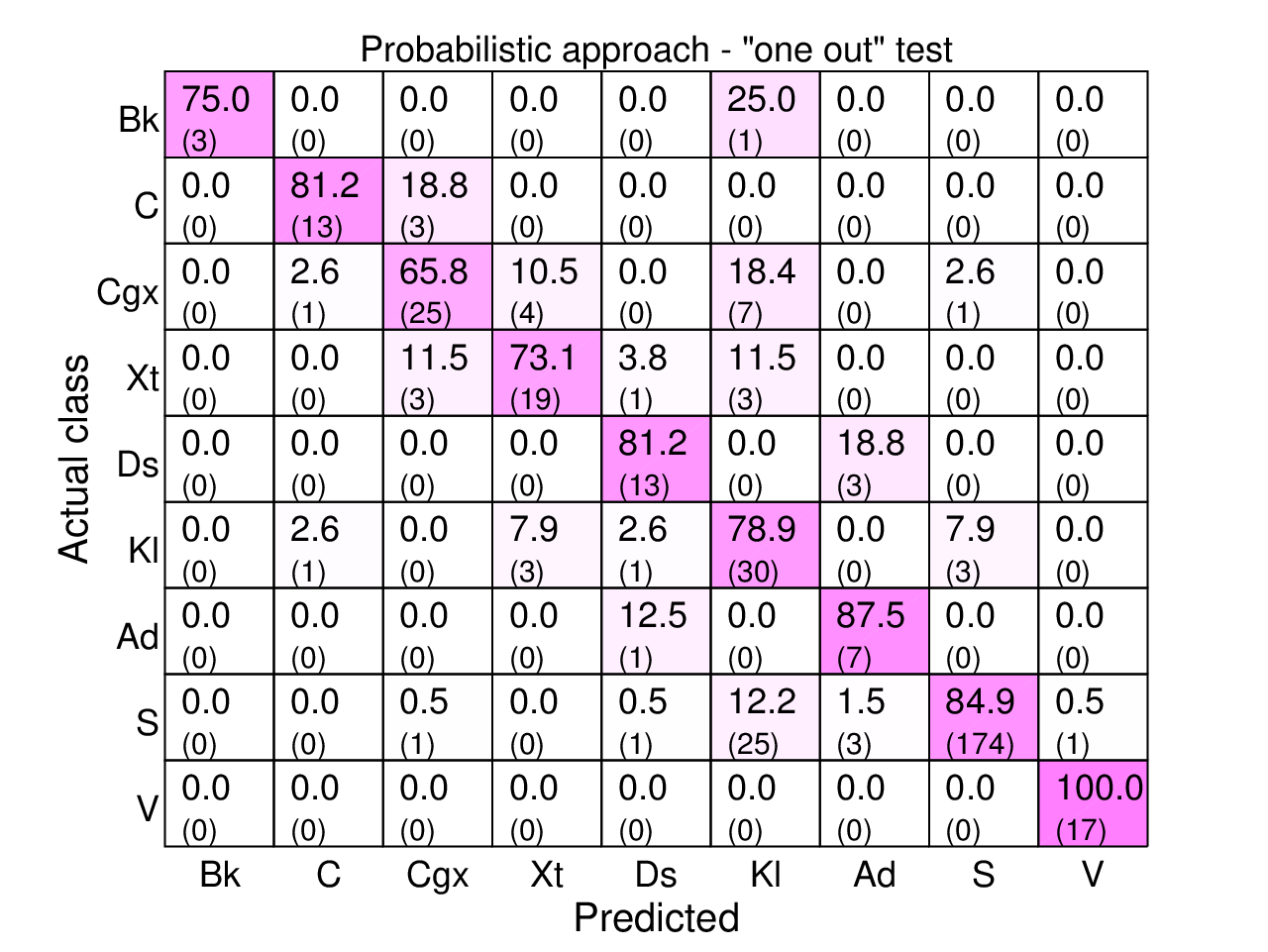}
\includegraphics[width=9.5cm]{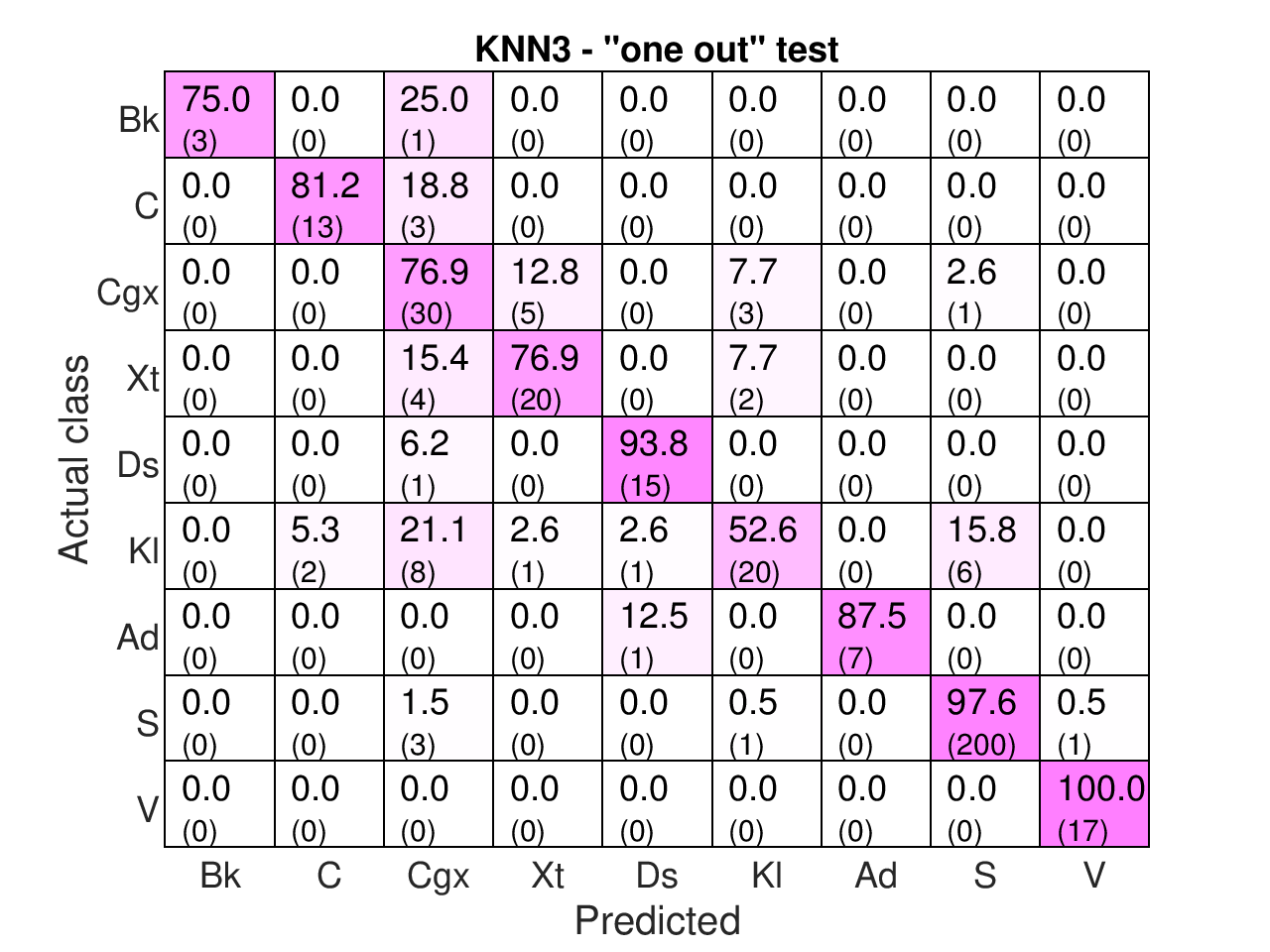}
\end{center}
\caption{The confusion matrix for the algorithms obtained after the "leave-one-out" cross-validation. It shows the classes predicted by the algorithm based on their computed colors versus the spectral classification (labeled actual class) for the 371 samples of the reference set. The top panel shows the results  for the probabilistic approach and the bottom panel shows the results for the k =3 nearest neighbors (KNN3) algorithm. Each row of the matrix represents the instances (in terms of percentages computed relative to the total number of objects for the given spectral class) in the actual spectral class while each column represents the instances in the class predicted  by the algorithm (the number of instances is given within the parenthesis). The color intensity outlines the interferences.}
\label{OneOut}
\end{figure}

The confusion matrix (i.e. a matrix layout that allows the visualization of the performance of the algorithm) obtained after "leave-one-out cross-test"  is shown in Fig.~\ref{OneOut} for both algorithms. This approach shows a high accuracy for the large groups (S-complex and C-complex). The accuracy for the end-members groups $B^{ni}$ and $A_d^{ni}$ which are defined on a very low number of objects, must be taken with caution. One of the most significant interferences shown by the graph is between the $D_s^{ni}$ and $A_d^{ni}$ types: both of them represent very red spectra and their near-infrared colors may coincide. 

In general the 3-color case derives more accurate results based on the fact that the increased number of degrees of freedom
puts additional constraints on the classification problem. The difference between the 3-color case and 2-color case is in the order of few percents, and it shows that the most relevant information is contained by the (Y-J) and (J-Ks) colors. 

Both algorithms report a classification for each of the 18\,265 asteroids which have the (Y-J) and (J-Ks) colors measured. The color errors are directly reflected into the probabilities of each class. If an object has an error bar that spans multiple classes it will have similar low probabilities for each of them. Moreover, it is likely that the two methods will provide different results preventing from concluding to a classification. This is also true for the objects which have color values at the border of different classes.

We recall that individual objects follow the statistics shown in (Fig.~\ref{OneOut}) and their classification is made with a probability which depends on their color errors. The tables accompanying the paper provide the information for each object: the colors (including the colors errors, and the time interval in which they were obtained), the classification results with the corresponding probabilities for each method and the final classification.

\section{Results}

We assigned a final taxonomic type based on KNN and probabilistic algorithms. This assignment was made when both methods gave the same result and the color errors are within the Q2 limits, i.e., (Y-J)$_{err} \leq 0.118$ and (J-Ks)$_{err} \leq 0.136$. It is the case of  6\,496 objects. All other,  which fail these conditions are marked with the letter \emph{U} (undefined) -- asteroids for which a final classification can not be made based on our dataset. In order to assign a final taxonomic classification, we did not constrained the probabilities to be higher than a certain threshold. However, we note that $75\%$ (and $90\%$ when considering only the 3NN probabilities) objects with a final taxon have both probabilities larger than 0.5 and the statistical results reported bellow are not sensitive to this selection.

The set of 6\,496 solar system bodies for which a final class was assigned includes: 144  $B_k^{ni}$, 613 $C^{ni}$, 197 $C_{gx}^{ni}$, 91 $X_t^{ni}$, 440 $D_s^{ni}$, 665 $K_l^{ni}$, 233 $A_d^{ni}$, 3\,315 $S^{ni}$, and  798 $V^{ni}$. These numbers are strongly dependent on the observational biases (high albedo objects are brighter and thus are more likely to be observed than the low albedo ones), and on the classification method (objects with prominent spectral features, i.e., V-types, are easily identified compared to those that show small NIR color variations between different types).

We do not assign a taxonomic type for about one third of the Q2 samples due to the fact that the results provided by the two algorithms (probabilistic and KNN3) are different. Some of the cases belong to color space domains where different groups overlap. Other unclassified asteroids may have compositions unaccounted by taxonomy, a point which is addressed in the discussion section using the (Y-J) versus (J-Ks) color plot.

Part of the asteroids reported in MOVIS-C catalog were classified using SDSS data \citep{2010A&A...510A..43C} and  have the visual albedo ($p_V$) determined based on WISE observations \citep{2011ApJ...741...68M, 2011ApJ...743..156M, 2014ApJ...792...30M}. Our results are discussed in the context of the information provided by these two large surveys. We used the Minor Planet Physical Properties Catalogue$\footnote{\url{https://mp3c.oca.eu/}}$ (MP$^3$C) and the Small Bodies Node of the NASA Planetary Data System\footnote{\url{https://pds.nasa.gov/}} (PDS) to retrieve the corresponding information.

The MOVIS-C catalog contains information for 6\,299 asteroids belonging to collisional families \citep{2015aste.book..297N}. Here we report a final taxonomic classification for 1\,874 (about $28.8\%$ from the classified objects) of them and we show how many asteroids of each class are assigned to families.  An in-depth study of NIR colors of asteroid families is provided by \citet{Morate2018}. 

\subsection{Comparison with the SDSS dataset}

In the sample of asteroids classified based on MOVIS-C NIR data, there are 1\,467 objects for which a taxonomic type was reported by \citet{2010A&A...510A..43C} using SDSS visible colors. This gives the opportunity to study the variation of colors over the visible to near-infrared interval for a large dataset.

\begin{figure}
\begin{center}
\includegraphics[width=9cm]{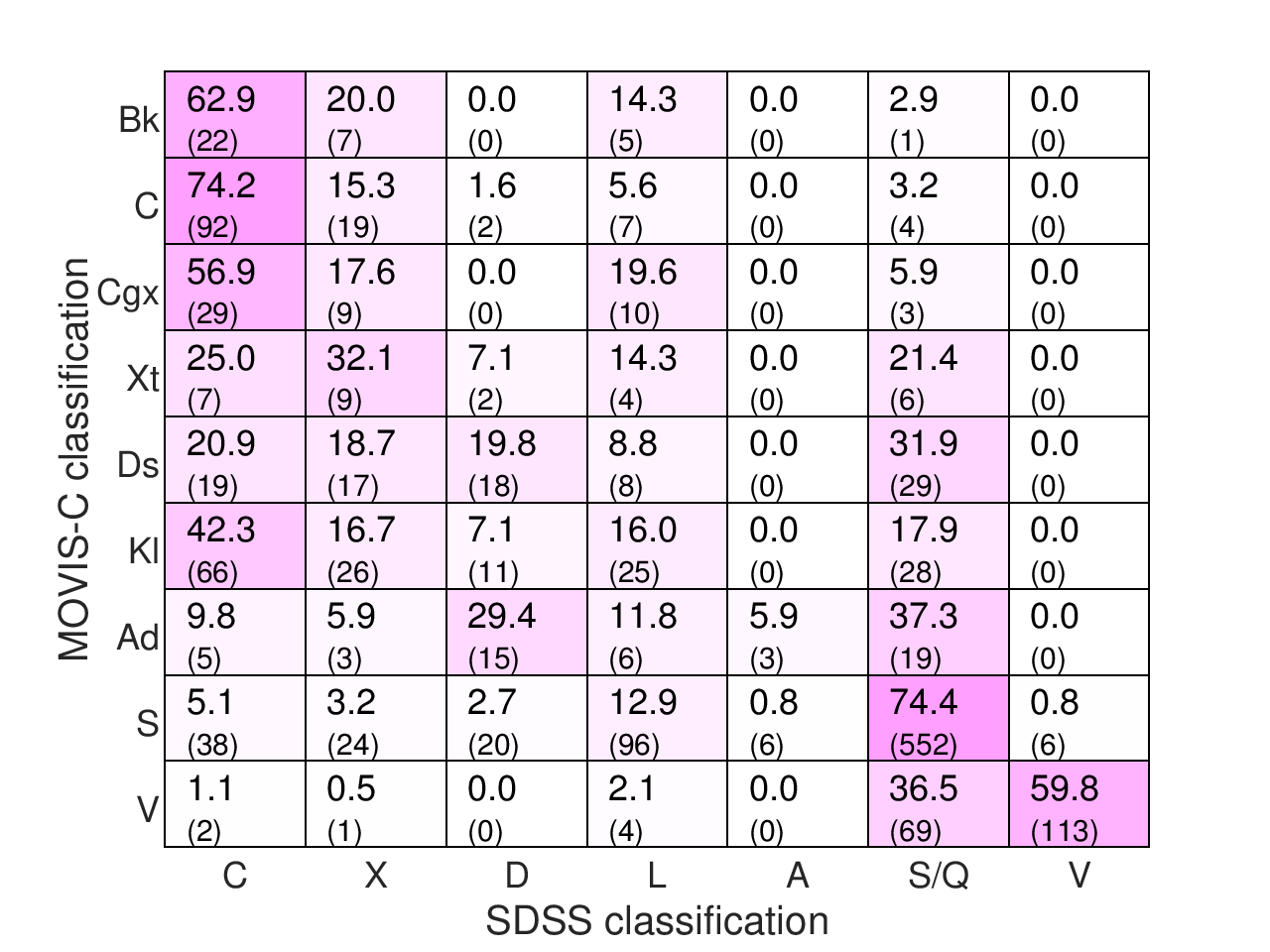}
\end{center}
\caption{The cross-tabulation matrix between the classification obtained based on MOVIS-C data (this work) and that obtained from SDSS data \citep{2010A&A...510A..43C}. The percentages are specified relative to MOVIS-C assigned type, and the number of objects is specified within parenthesis. The color intensity outlines the interferences.}
\label{SDSSandVIS}
\end{figure}

Following the definitions of the taxonomic types from this paper and those from \cite{2010A&A...510A..43C} we computed the cross tabulation matrix for the two classifications (Fig.~\ref{SDSSandVIS}). At first glance this shows the matching between the two results. Furthermore, this cross matrix also underlines spectro-photometric behavior not accounted by the current taxonomies. 

There is a significant matching between the two sets of data for identifying carbonaceous (i.e. objects characterized by featureless spectra) and olivine-pyroxene (i.e. objects characterized by bands at 1 and 2 $\mu$m) compositions. The main parts of the cross tabulation matrix correspond to ($B_k^{ni}$, $C^{ni}$, $C_{gx}^{ni}$, $X_t^{ni}$)$_{MOVIS}$ vs (C,X)$_{SDSS}$ -- upper left, and ($S^{ni}$, $V^{ni}$)$_{MOVIS}$  vs (S/Q,V)$_{SDSS}$ -- lower right. They outline the main compositional types.

About $68.1 \%$ (143 asteroids) from the objects classified as $B_k^{ni}$, $C^{ni}$, $C_{gx}^{ni}$ correspond to C$_{SDSS}$ group, while a fraction of $16.7\%$ of them are found as X$_{SDSS}$. Concerning the silicate-like compositions, there are 552 objects belonging to S-complex according to both data sets ($78.4 \%$ matching).  Some of the $S^{ni}$ asteroids are reported as L$_{SDSS}$-type ($13.6\%$, 96 asteroids).

Although the "leave one-out test" gives a probability higher than 80$\%$ for classifying the $D_s^{ni}$-types, there is a significant difference when comparing them with their classification based on SDSS  data (Fig.~\ref{SDSSandVIS}). An explanation for the 29 objects reported as $D_s^{ni}$ and as $S_{SDSS}$ are the asteroids with high content of olivine which show a red spectrum and a weak 2 $\mu m$ band, like those classified as Sa (Fig.~\ref{ReflectanceClasses}) by \citet{2009Icar..202..160D}. Also the S-type objects with high NIR spectral slopes (marked with "w" in Bus-DeMeo taxonomy) may account as an explanation for the miss-match. The albedo values provided by the WISE survey confirm the silicate nature of this sample.

Only 25 objects are found as $K_l^{ni}$ - $L_{SDSS}$. Most of the $K_l^{ni}$ asteroids were classified by \citet{2010A&A...510A..43C} as $C_{SDSS}$-types (66 objects - $42.3\%$) or $S_{SDSS}$-types (28 objects - $17.9\%$). This is due to the fact that K and L types (as were defined by Bus-DeMeo taxonomy), do not show a prominent spectral features in the NIR - typical of the S-types, but neither they are featureless, like the C-types. 

There are only three olivine rich asteroids (described by $A_d^{ni}$ group) identified by both surveys. About 19 objects classified as $A_d^{ni}$ are found as $S_{SDSS}$-type. The NIR colors suggest for this sample an olivine rich composition or may indicate a reddening due to space-weathering effect. A noticeable interference consists of the 15 asteroids ($29.4\%$) found as $A_d^{ni}$ and which were reported as $D_{SDSS}$. This crosstalk is also seen on the "leave one out" test (Fig.~\ref{OneOut}). Due to the reference set, which does not have D-types with high NIR spectral slope, all MOVIS-C data with extremely red NIR colors are reported as $A_d^{ni}$. In this case albedos, or SDSS colors, provide a way to discriminate between the D- and A- types from Bus-DeMeo taxonomy.

A particular case is that of 24 asteroids with SDSS C-type colors and found as  $A_d^{ni}$ or $D_s^{ni}$. This suggests a spectral behavior neutral or slightly red in the visible which turns red to extremely red in the NIR wavelengths. A similar behavior has been previously reported by \citet{2012Icar..218..196D}, showing that asteroids classified as B-types from their visible spectra presented a variety of shapes in the near-infrared, from blue to moderately red spectral slopes.

\begin{figure*}
\begin{center}
\includegraphics[width=5cm]{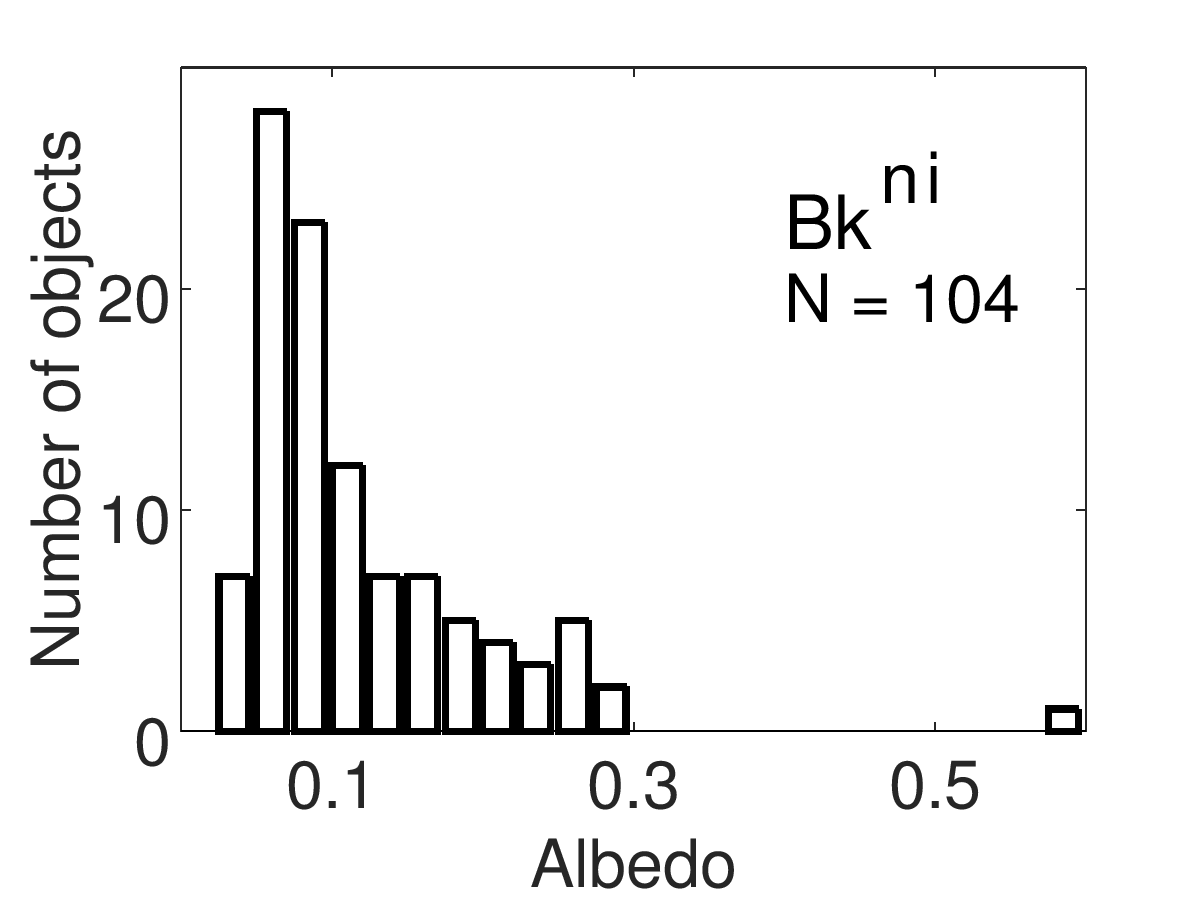}
\includegraphics[width=5cm]{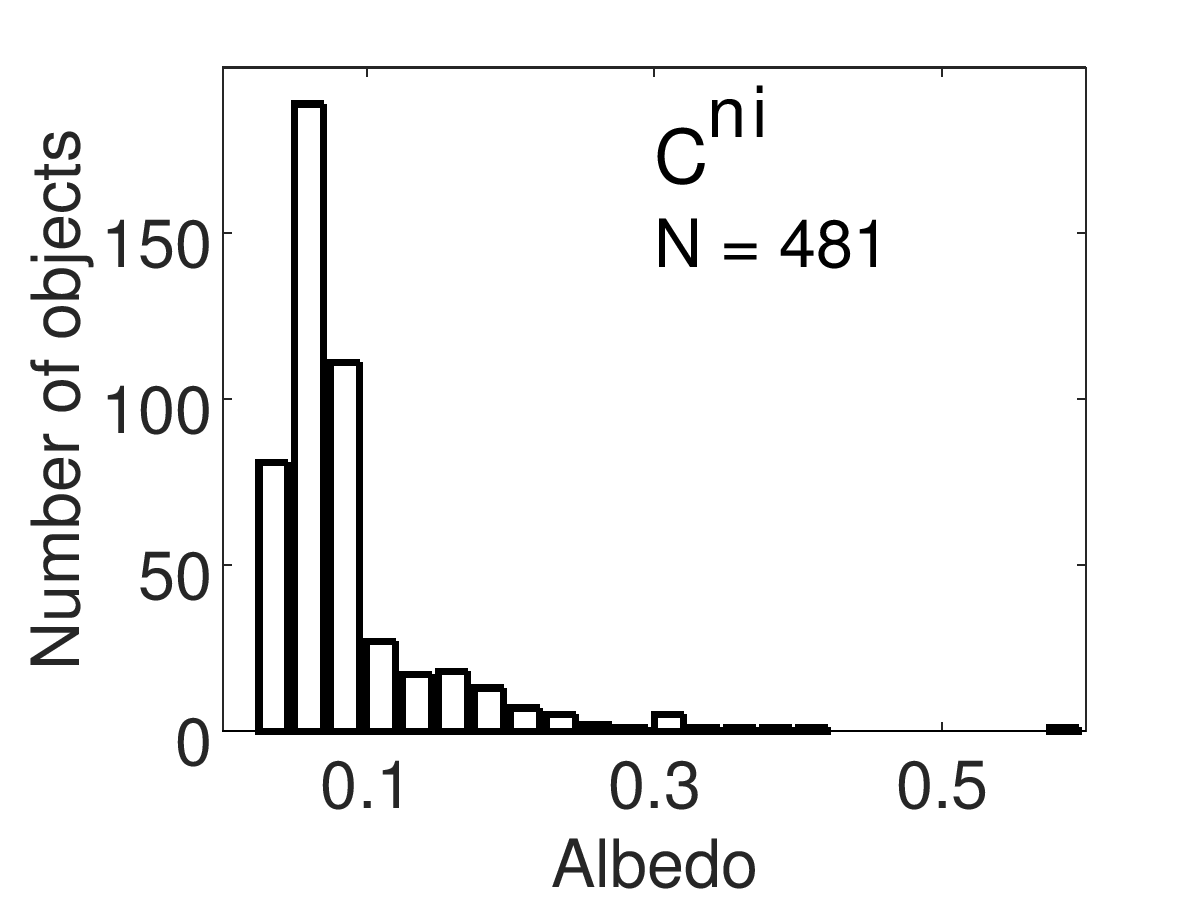}
\includegraphics[width=5cm]{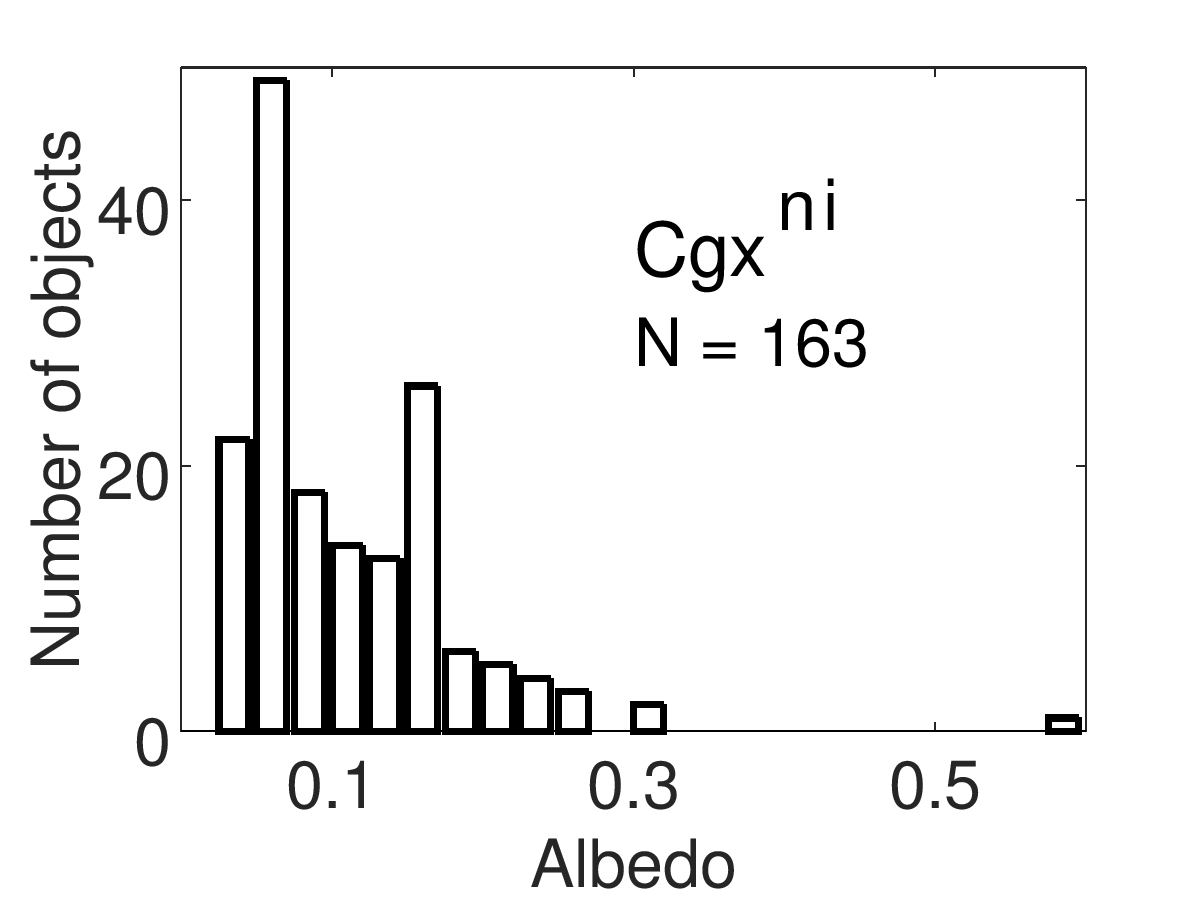}
\includegraphics[width=5cm]{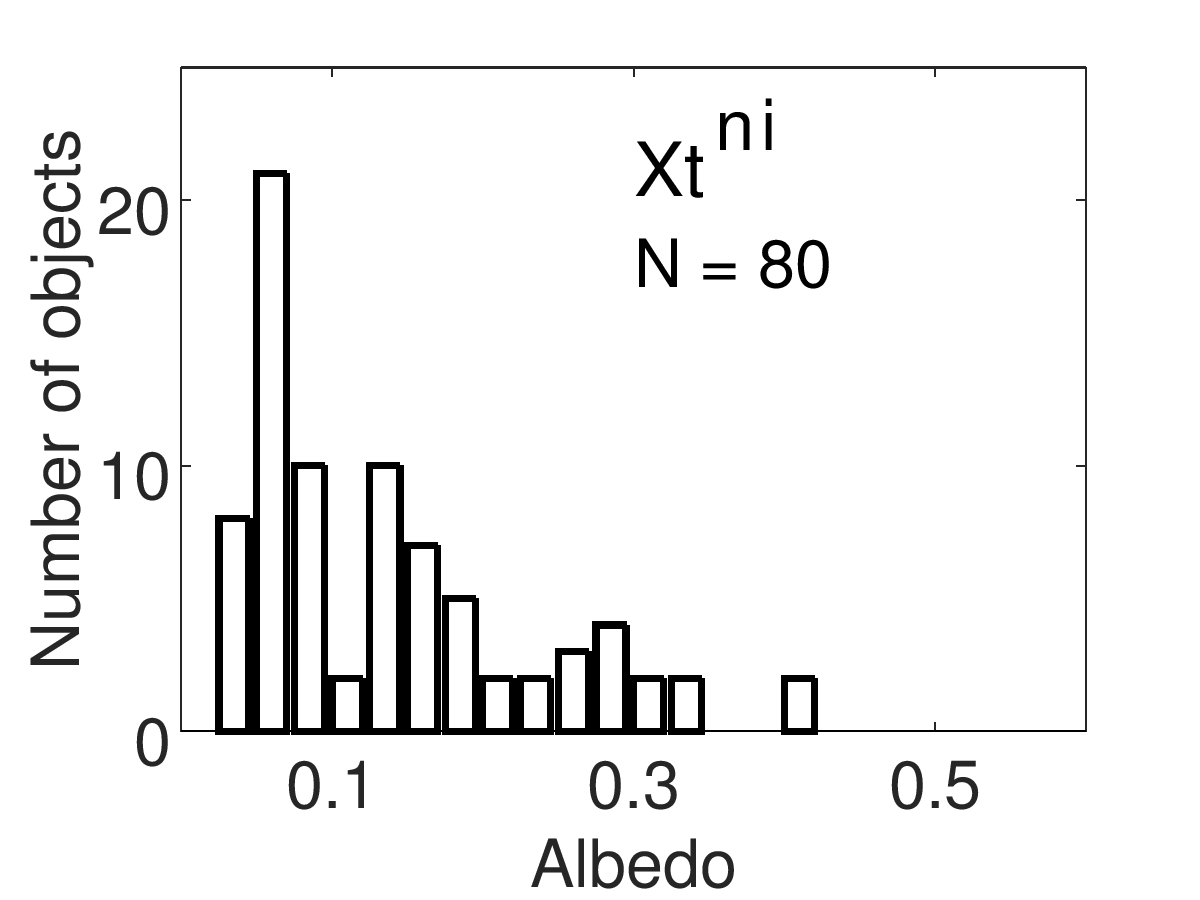}
\includegraphics[width=5cm]{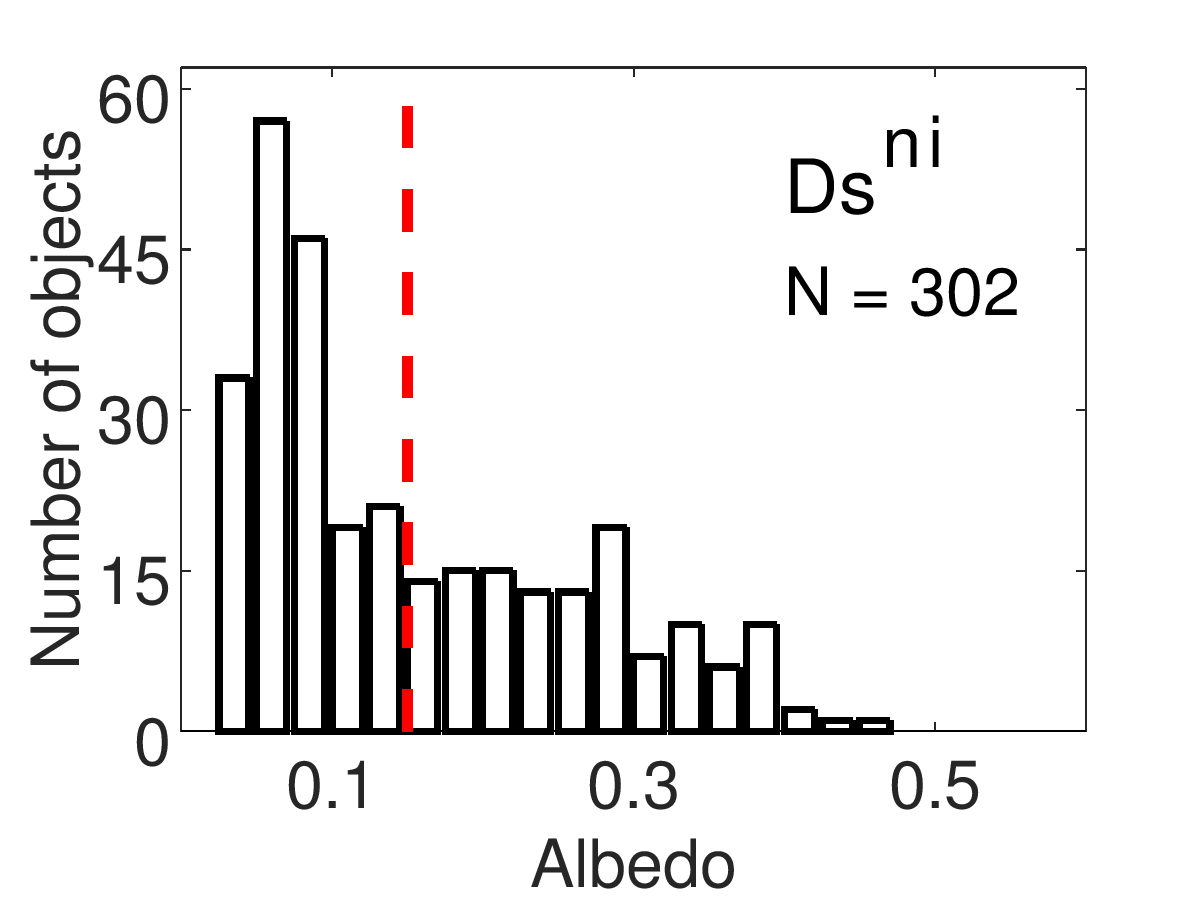}
\includegraphics[width=5cm]{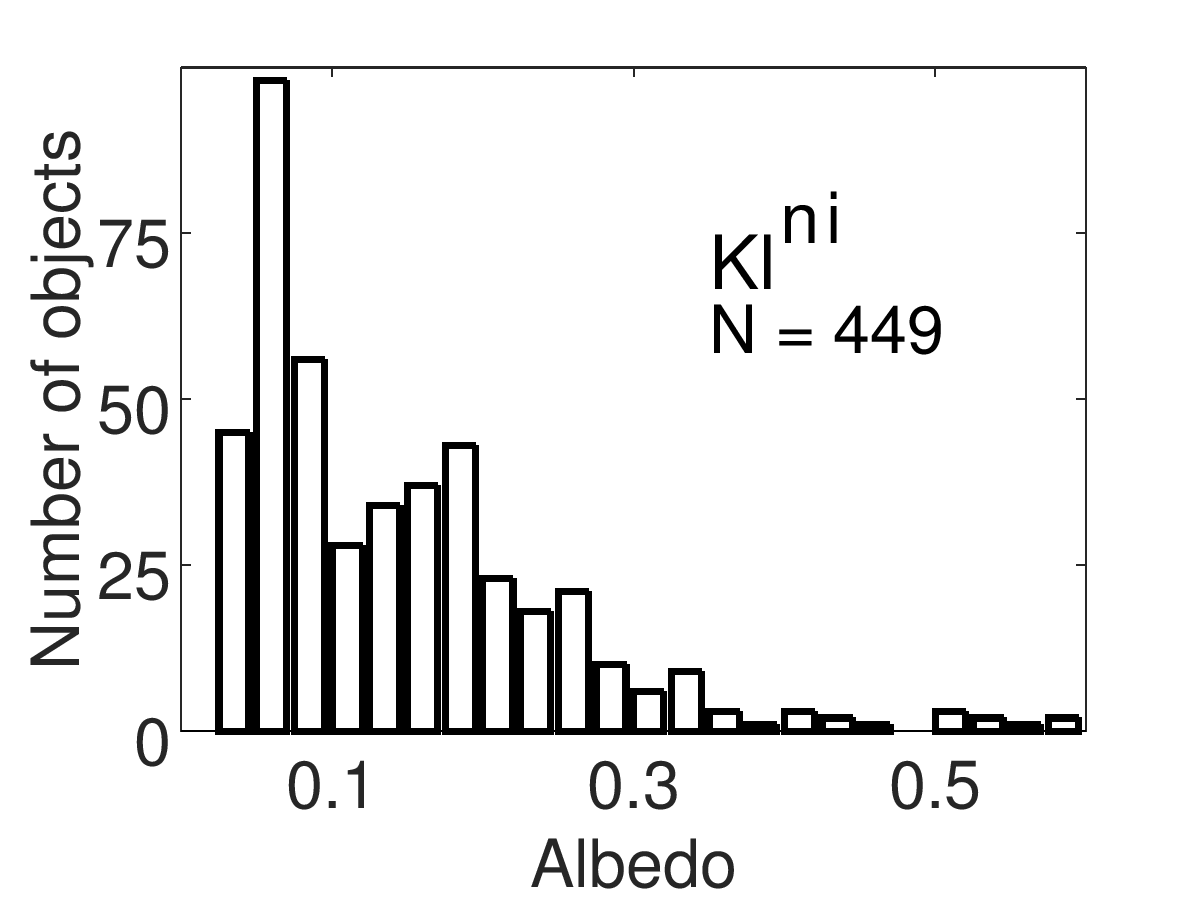}
\includegraphics[width=5cm]{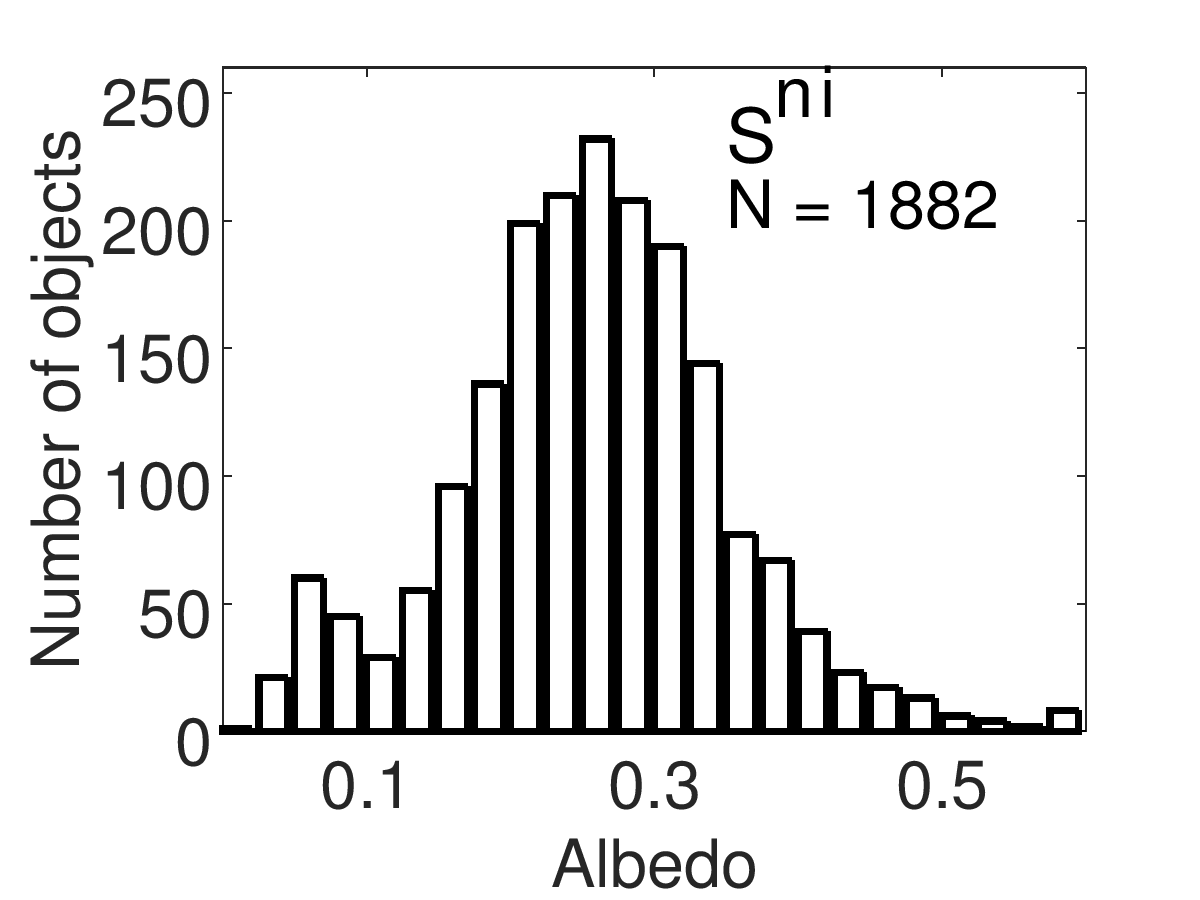}
\includegraphics[width=5cm]{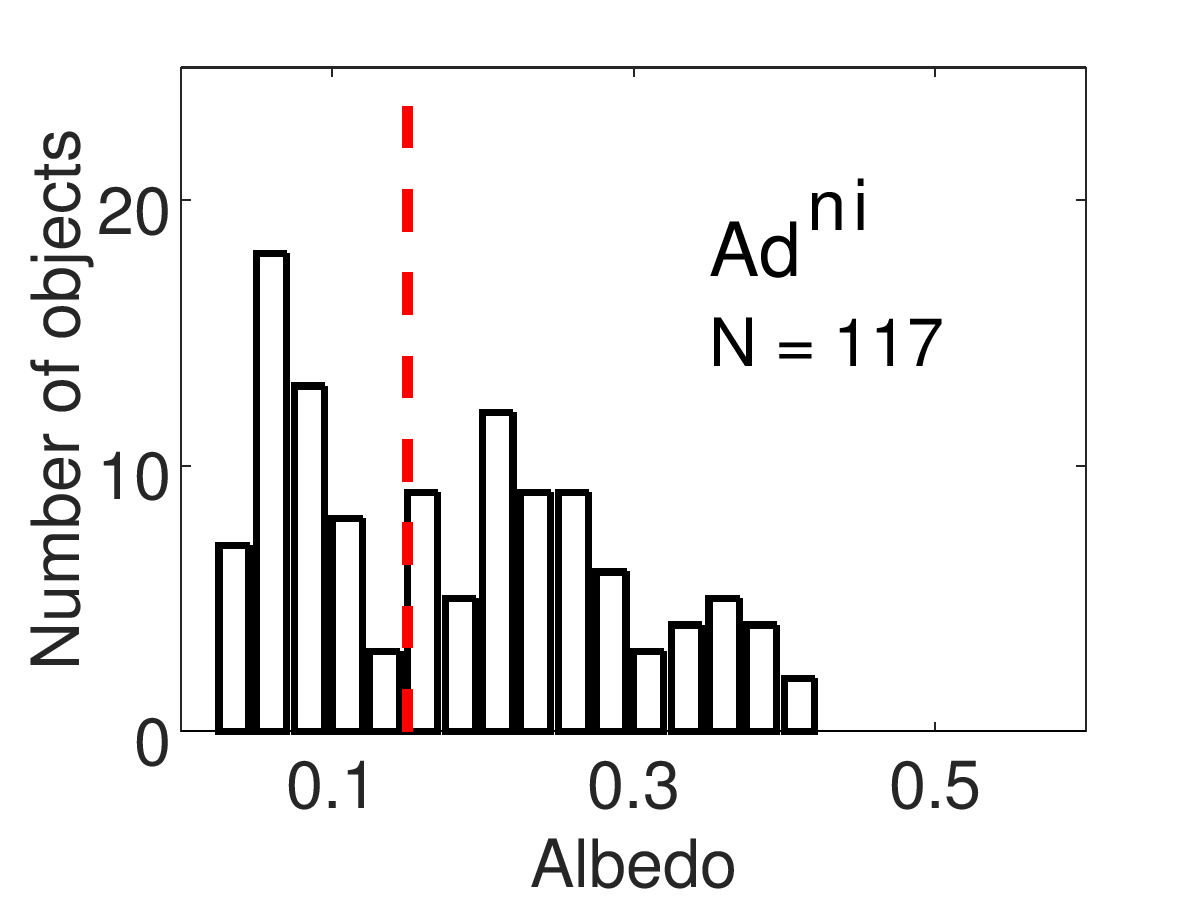}
\includegraphics[width=5cm]{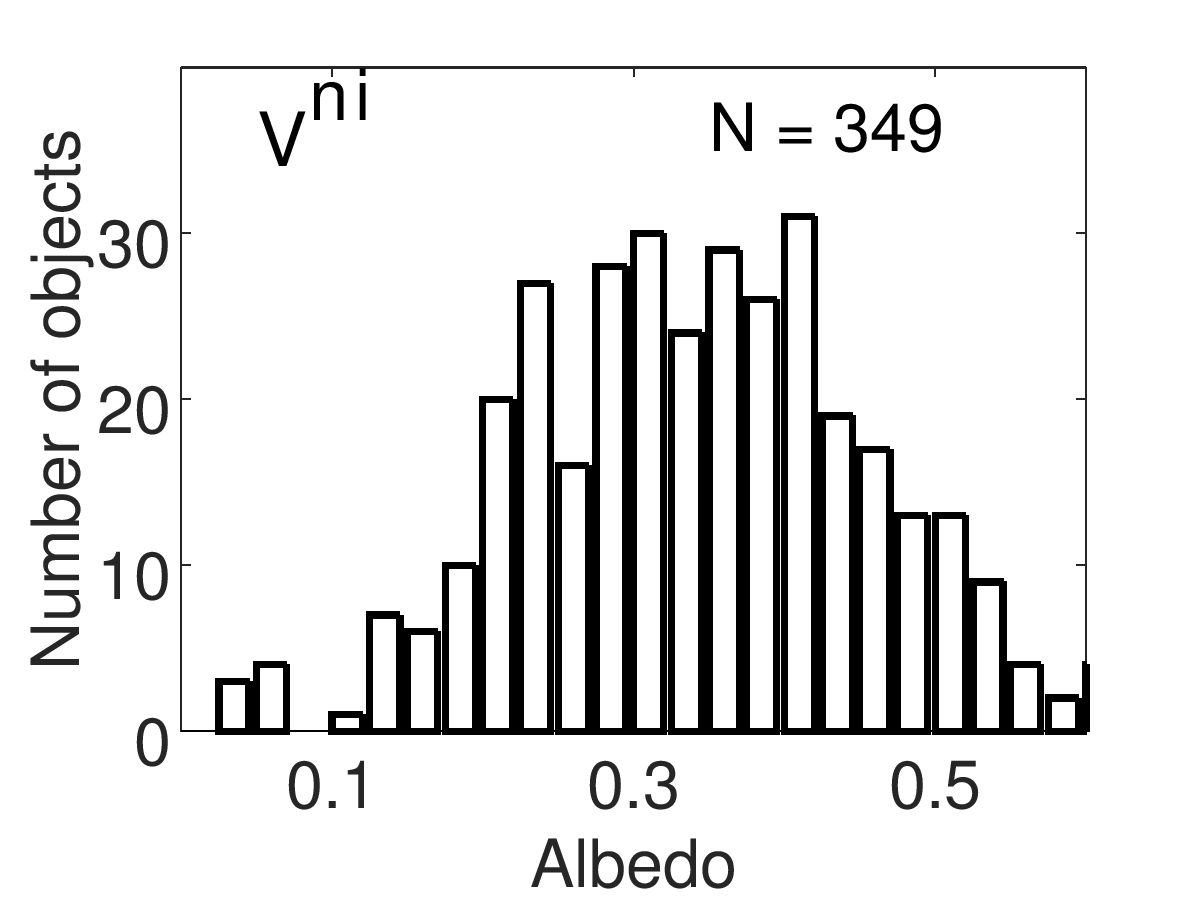}
\end{center}
\caption{Visible albedo distributions of asteroids classified based on MOVIS-C data. The bin size is 0.025. Each histogram shows the taxonomic group (defined in Section 2.2) and the number of objects (N) used. The vertical red bars mark the $p_V =0.15$ limit used for the discussion of $A_d^{ni}$ and $D_s^{ni}$ types.
}
\label{TypesAlbedo}
\end{figure*}

We note that spectral studies have to confirm the characterization of individual objects because the color errors reported by both surveys can significantly change the conclusion. The Fig.~\ref{SDSSandVIS} provides a statistical view of the cross-matching between the two large datasets. The relevance of the percentages has to consider the number of instances (i.e. the number of objects reported for each element of the matrix).

\subsection{Extremely red colors: $A_d^{ni}$ and $D_s^{ni}$}

The asteroids classified as $A_d^{ni}$  have (J-Ks)$\gtrsim0.65$ and the average value (J-Ks)$_{avg} = 0.84\pm 0.17$. This is the only group with such a red color (see Fig.~\ref{CompColor371Spectra} and Table~\ref{TaxonColor}). The distribution of their albedo, available from the WISE measurements for 117 of these objects, spans large intervals covering both silicaceous and carbonaceous compositions (Fig.~\ref{TypesAlbedo}). 

A-type asteroids \citep[e.g.][]{1983Icar...55..177V,2002Icar..158..146B,2009Icar..202..160D} are associated with olivine dominated compositions \citep[e.g.][]{2014Icar..228..288S}. The median albedo of the five A-type objects considered by \citet{2011ApJ...741...90M} is $p_V =  0.191\pm0.034$. Thus, the histogram of albedo (Fig.~\ref{TypesAlbedo}) shows that the group we defined as $A_d^{ni}$ contains not only olivine rich objects but also low albedo ones. We can assume that the low albedo asteroids identified as $A_d^{ni}$ are very red D-types.

By examining Fig.~\ref{TypesAlbedo} it is reasonable to assume a limit of $p_V=0.15$ for separation of these two groups. As a result, there are about high albedo asteroids that can be associated with olivine dominated compositions. About 23 of them were observed by SDSS and 22 are reported as A/L/S (A-3, L-4, S-15) by \citet{2010A&A...510A..43C}, confirming their red spectra in the visible region and the start of the 1 $\mu$m feature.

These high albedo $A_d^{ni}$ asteroids have a wide distribution of albedos with an average of $p_V = 0.26\pm0.07$. With respect to their semi-major axis ($a$),  most of them are in the inner part of the main belt ($2.0\leq a <2.5$ au )- about $43\%$, and in the middle part of the main belt ($2.5\leq a <2.82$ au) - about $41\%$. About $16\%$ of high albedo $A_d^{ni}$  are in the outer part of the main belt ($2.82\leq a <3.2$ au).

Only 11 out of this sample of 68 objects are associated with some families by \citet{2015aste.book..297N}. This fraction is unexpectedly low, as one way of forming olivine-rich composition is through magmatic differentiation, being the major constituent of the mantles of most differentiated bodies \citep[e.g.][]{1996M&PS...31..607B, 2004A&A...422L..59D, 2007M&PS...42..155S, 2014Icar..228..288S}. Thus, we may expect to find them among the members of collisional families as a result of disruption of differentiated bodies.

The sample of low-albedo $A_d^{ni}$ comprises 49 objects. They have an average albedo value of $p_V = 0.08\pm0.03$ which suggests that they are very red D-types. The classification of \citet{2010A&A...510A..43C} confirms their assignment as D-types (10/13 objects). Most of these asteroids are in the outer part of the main belt - 27 ($55\%$), while some of them are in {bf the Cybele and Hilda populations}. Five of these objects with extremely red color and low albedo are found in the inner part of the main belt. 

The red color of these presumed D-types, and their low albedos values suggest a possible abundant content in organics and volatiles. Such compositions are found among objects that are considered the most primitive ones in the solar system. They are identified in the Jupiter Trojan region and in the Kuiper Belt. \cite{2009Natur.460..364L} hypothesized that they had formed in the outer solar system and had been captured in their present locations as a consequence of the migration of the giant planets. 

The equivalent colors of spectra classified by \citet{2009Icar..202..160D} as D-types are less red, $<(J-Ks)>$~= 0.593$\pm$0.074, than our low albedo $A_d^{ni}$ candidates. Spectrally, the  D-types are defined by their high slope and lack of features in the visible and near-infrared wavelength range. We found 440  $D_s^{ni}$ candidates based on NIR colors. The albedo distribution made for 302 of them (Fig.~\ref{TypesAlbedo}), shows that this group is not well constrained, having both low and high albedo objects. This might be explained with some of the reddest S-complex asteroids having similar color values as those of D-types.

Following the same approach as for $A_d^{ni}$, we assumed a limit of $p_V = 0.15$. This provides 176 low-albedo $D_s^{ni}$ from 302 candidates with measured albedo. The selected sample has a narrow distribution of albedo ($p_V = 0.08\pm0.03$), and spans over a large range of sizes (2 -- 100 km). The average value of albedo is higher than the one reported by \citet{2011ApJ...741...90M} who found a $p_V = 0.048\pm0.025$ for 13 D-types spectroscopically classified. The selection bias implies that objects with high albedo are more likely to be observed than darker objects and may account for the differences. The majority of them ($67\%$) are located in the outer part of the main belt and in the Cybele, Hilda and Trojan populations. 

A fraction of $10\%$ low-albedo $D_s^{ni}$ are inner main belt asteroids. This is an uncommon location for these type of objects. Such interlopers were identified also based on the SDSS data and confirmed with follow-up observations by \citet{2014Icar..229..392D}. They proposed various scenarios for their possible origin, varying from a distinct compositional group to different migration mechanisms. With the exception of (1715) Salli, all our inner main belt D-type candidates have a diameter in the range of 2-6 km, supporting the hypothesis of Yarkovsky drift across the resonances which is more efficient for small asteroids \citep{2002aste.book..395B,Bottke:2006aa}.

In general ($\sim83\%$), our low-albedo D-type candidates are not associated with collisional families by \citet{2015aste.book..297N}. However, seven objects are associated with (24) Themis family and six objects associated with (221) Eos family.

\subsection{Blue colors - $B_k^{ni}$}

The asteroids with (Y-J)$\leq$0.219 (which is the value for the Sun) are classified as $B_k^{ni}$. In the Bus-DeMeo taxonomy, the B-types represent the blue end member class and are distinctive because of their negative spectral slopes.

Objects classified as B-type asteroids present low albedo values and are mostly found in the middle and outer part of the main belt. Using a sample of 22 asteroids classified as B-types from their visible spectra, \citet{2010JGRE..115.6005C} divided in three main groups, according to their NIR spectra: those with spectra similar to that of (2) Pallas; those similar to NIR spectrum of (24) Themis; and those not falling into either the Pallas or the Themis group. In a later publication, \citet{2012Icar..218..196D} showed that, instead of 3 groups, the NIR spectra of asteroids classified as B-type according to their visible spectra, presented a continuous shape variation from negative, blue spectral slopes to positive red slopes. This continuum in spectral slopes was also present in the sample of carbonaceous chondrites that best resembled the spectra of B-type asteroids in both papers. Objects having negative, blue spectral slopes in the NIR, i.e., those that will belong to our $B_k^{ni}$ considering their NIR colors, are associated with CV, CO, and CK carbonaceous chondrites. New findings about the composition of these objects are expected to be reveled by the  NASA's OSIRIS-Rex  mission whose primary target is the B-type near-Earth asteroid Bennu \citep{Lauretta:2017aa, 2010ApJ...721L..53C}.

The colors computed from the reference spectra show that $B_k^{ni}$ is a distinct group in the (Y-J) vs. (J-Ks) color space. Based on these reference values we classified 144 asteroids in this category. The albedo was measured for 104 objects of this sample, and their histogram shows two groups (Fig.~\ref{TypesAlbedo}). The largest group has a narrow distribution and a peak around $p_V = 0.075\pm0.022$. The second group presents a wide range of albedo values, from $\sim$0.12 to 0.30, with a shallow peak around $p_V\approx0.15$. For comparison, \cite{2011ApJ...741...90M} found a median value of $p_V = 0.120\pm0.022$ for the B-types with V-NIR spectral data (we note that they used only two objects). Our finding does not confirm this value - there is no peak in the albedo distribution corresponding to it. A detailed study of albedo values for asteroids classified as B-types was performed by \cite{Ali-Lagoa:2013aa}. They found an average value of $p_V = 0.07\pm0.03$ which is in agreement with our result.
 
Less than half of our $B_k^{ni}$ sample (about 61 asteroids) are associated with asteroid families. Among them, there are 6 objects belonging to (10) Hygiea, 9 to (24) Themis, and 8 to (668) Dora. All of these correspond to our low albedo group of $B_k^{ni}$. 

The largest fraction of $B_k^{ni}$ associated with collisional families is of 24 objects linked to (221) Eos. This is one of the outer main belt families that has over 9\,000 members and it is recognized as K-type \citep{2015aste.book..297N}. \citet{2008Icar..195..277M} noted spectral similarities between the Eos family asteroids and CO3, CV3, and CK carbonaceous chondrites. The albedo values are available for 16 out of these 25 objects, and its average value is $p_V = 0.166\pm0.042$. This is in agreement with the value of  $p_V = 0.163 \pm 0.035$ reported by \citet{Masiero:2015aa} for the Eos family.

Although it is not outlined by the colors computed from reference spectra, the albedo values and the link between some objects and the family of (221) Eos show a marginal mixture of $B_k^{ni}$ with possible spectrally K-types which have blue NIR colors \citep{Morate2018}.

\subsection{The C- and X- complexes: $C^{ni}$, $C_{gx}^{ni}$ and $X_t^{ni}$ }

Even though they are spectrally distinctive, the types corresponding to the C- and X- complexes span similar intervals of NIR colors values. By considering the broad compositional types, we clustered these two complexes in three groups. The $C^{ni}$ and $C_{gx}^{ni}$ are separated mostly by the (J-Ks) with a difference less than $\sim$ 0.05 mag. The $X_t^{ni}$ shows a redder (Y-J) compared with the other two groups. It has to be noted that these groups show a uniform distribution over the region they occupy in the color space.

With respect to these groups, the largest number of asteroids (a total of 613) was assigned to the $C^{ni}$ group. Their albedo distribution (available for 481 out of these 613 objects), proved that the separation was successful. A fraction of $\sim80\%$ are low albedo objects ($p_V\leq0.10$) compatible with a carbonaceous like composition. About $41\%$ of asteroids classified in this group (196 objects) are linked to asteroid families. Among them, the largest fraction of objects belongs to (10) Hygiea, and to (24) Themis.

All three groups associated with the C/X complex show a double peak distribution of albedo (Fig.~\ref{TypesAlbedo}). A first peak is centered around $p_V = 0.06\pm0.02$ and a second one at $p_V \sim 0.15\pm0.03$. This last one is more pronounced for $C_{gx}^{ni}$ and $X_t^{ni}$ and can be associated with metallic compositions \citep{2010Icar..210..655F, 2010M&PS...45..304C}. We note that in Tholen taxonomy \citep{1984PhDT.........3T} the X-complex is divided in three categories: P - "primitives" (low albedo objects), M - "metallic" (medium albedo ones, i.e $\sim 0.15$), and E - "enstatites" (high albedo asteroids). 
\subsection{The $K_l^{ni}$ group}

In the NIR color-color plots the $K_l^{ni}$ group lies in between the S-, C-, and X- complexes. The 665 asteroids classified as $K_l^{ni}$ include many miss-identifications from these three large complexes. This is outlined by their albedo distribution (Fig.~\ref{TypesAlbedo}).

We note that about half {(98 out of 214)} of the asteroids classified in this group and associated with asteroid families are linked to (221) Eos (recognized as K-type family). We showed that some of the (221) Eos family members have blue colors (thus were classified as $B_k^{ni}$), but the albedo is compatible with K-type. This suggests that the group covers a larger color interval than expected from the reference set.

\subsection{The S complex asteroids: $S^{ni}$}

Asteroids classified as $S^{ni}$ are spread over a broad region in the (Y-J),(J-Ks) and (H-Ks) color space. They cover a continuous interval of NIR colors. Using a sample of objects spectrally classified by SMASS \citep{1995Icar..115....1X,2002Icar..158..106B} and S3OS2 \citep{2004Icar..172..179L}, \citet{2016A&A...591A.115P} showed that S-complex asteroids are clearly separated in the (Y-J) vs (J-Ks) plot, forming a distinct cluster with a continuous distribution.

In this work we used the reference spectra of S, Sq, Sr, Sv and Q-types from \citet{2009Icar..202..160D} to define our $S^{ni}$ group (note that Sa was included in the $A_d^{ni}$ group). A total of 3\,315 asteroids are assigned to this group by the two algorithms. It represents more than half of the sample for which we report a taxonomic classification.

WISE albedos are available for 1\,882 asteroids classified as $S^{ni}$. They show a Gaussian distribution with a peak at $p_V = 0.26\pm0.10$. This is close to the median value of $p_V=0.223\pm0.073$ reported by \citet{2011ApJ...741...90M} for asteroids belonging to the S-complex according to the Bus-DeMeo taxonomy. A small fraction of asteroids ($\sim4\%$) have albedo values lower than 0.10 (outliers), and are most likely miss-identified D-types, due to the proximity of the two groups in the color-color plot (Fig. \ref{CompColor371Spectra}). 

We found asteroids in the $S^{ni}$ group having a variety of orbital parameters, from near-Earth asteroids to outer main belt populations. The few objects identified in the Hilda and Trojan populations present albedo values that do not confirm a silicate-like composition and are most likely miss-identifications of D-types. A notable exception is asteroid (3675) Kemstach which we classified as $S^{ni}$, and it has orbital elements that place it into Cybele population.  This is known to be dominated by primitive C and D types. Our classification of (3675) Kemstach  is in agreement with the albedo value of $p_V = 0.181\pm0.018$ and with the taxonomic type reported by SDSS. The visible spectra obtained by \citet{2005A&A...432..349L} confirm its classification as an S-complex object.

About $30\%$ (875 objects) of asteroids classified as $S^{ni}$ were associated with various asteroids families. In this sample, 251 out of 875 asteroids are associated with the (15) Eunomia family (a large middle main belt family dominated by the S-complex asteroids), with over 5\,000 members \citep{2015aste.book..297N}. 

We also note the presence of $S^{ni}$ asteroids in families dominated by very different compositions compared to S-types, like (4) Vesta, (93) Minerva, (135) Hertha, (158) Koronis, and (221) Eos. As noted by \citet{2017A&A...600A.126L} and \citet{Morate2018}, the Vesta family includes a non-negligible fraction of S-types ($\approx 11\%$).

\subsection{Basaltic asteroids - $V^{ni}$}

Asteroids with (Y-J)$\geq  0.5$ and (J-Ks)$\leq 0.3$ were identified by \citet{2017A&A...600A.126L} as V-types. They found  477 V-type candidates with color magnitude errors less than 0.1. A sample of 244 of these V-type candidates are not associated with the Vesta family. As the only confirmed source of basaltic material is (4) Vesta, several dynamical mechanisms were proposed to explain the large fraction of basaltic asteroids outside the family \citep{2014MNRAS.439.3168C, 2017MNRAS.468.1236B}.

Here we report 798 objects classified as $V^{ni}$ based on the updated version of the MOVIS catalog, and with a less restrictive limit for the error (Q2). About 694 objects follow the conditions of (Y-J)$\geq 0.5$ and (J-Ks)$\leq 0.3$. The $V^{ni}$ asteroids with colors outside these limits are in the vicinity of the S-complex region and lie in the continuous region in between these two groups.

The $V^{ni}$ group associated with basaltic asteroids shows the broadest distribution in color space, outlined in terms of mean and standard deviation values: (Y-J)$~= 0.632\pm0.122$, (J-Ks)$~= 0.060\pm0.139$, and (H-Ks)$~= -0.078\pm0.172$. \citet{2017A&A...600A.126L} noted that the (Y-J) colors of the Vesta family candidates seem to have a narrower distribution than the colors of the non-Vesta family asteroids. Visible and near-infrared spectral surveys \citep[e.g.][]{2016MNRAS.455.2871I,2017MNRAS.464.1718M,2018MNRAS.475..353M} of these spectro-photometric V-type candidates  are required in order to confirm how this color variation translates into actual compositional differences.
 
The albedo values of asteroids classified as $V^{ni}$ present a broad distribution (Fig.~\ref{TypesAlbedo}), centered around $p_V = 0.352\pm0.121$. This is similar to the median value of $p_V = 0.362\pm0.100$ reported by \citet{2011ApJ...741...90M} for asteroids classified as V-types according to the Bus-DeMeo taxonomy. Although being quite speculative, we cannot discard the possibility that the two additional peaks at $p_V = 0.30$ and $p_V = 0.23$ might be indicative of two different compositional sub-groups inside V-types. For an in depth analysis of V-type asteroids please refer to \citet{2017A&A...600A.126L}.

\begin{figure*}
\begin{center}
\includegraphics[width=9cm]{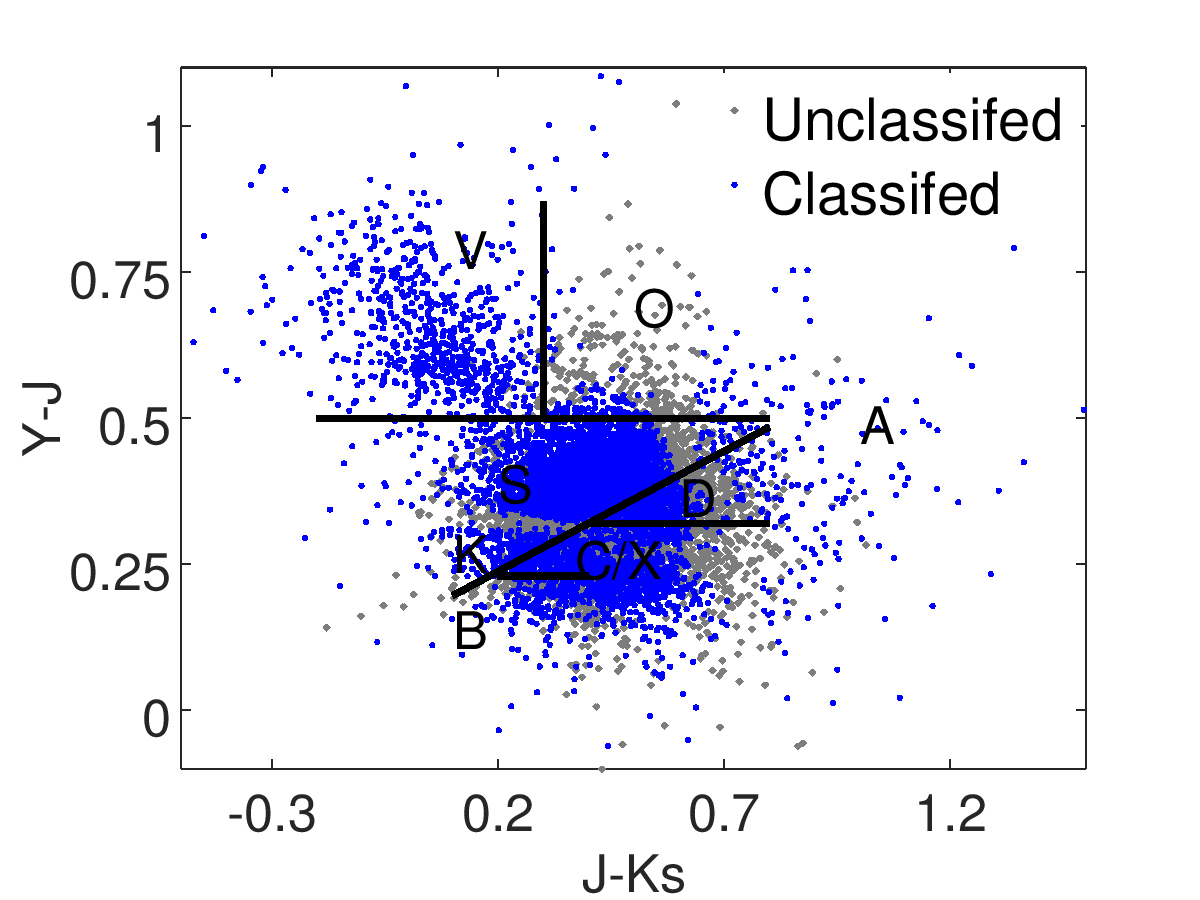}
\includegraphics[width=9cm]{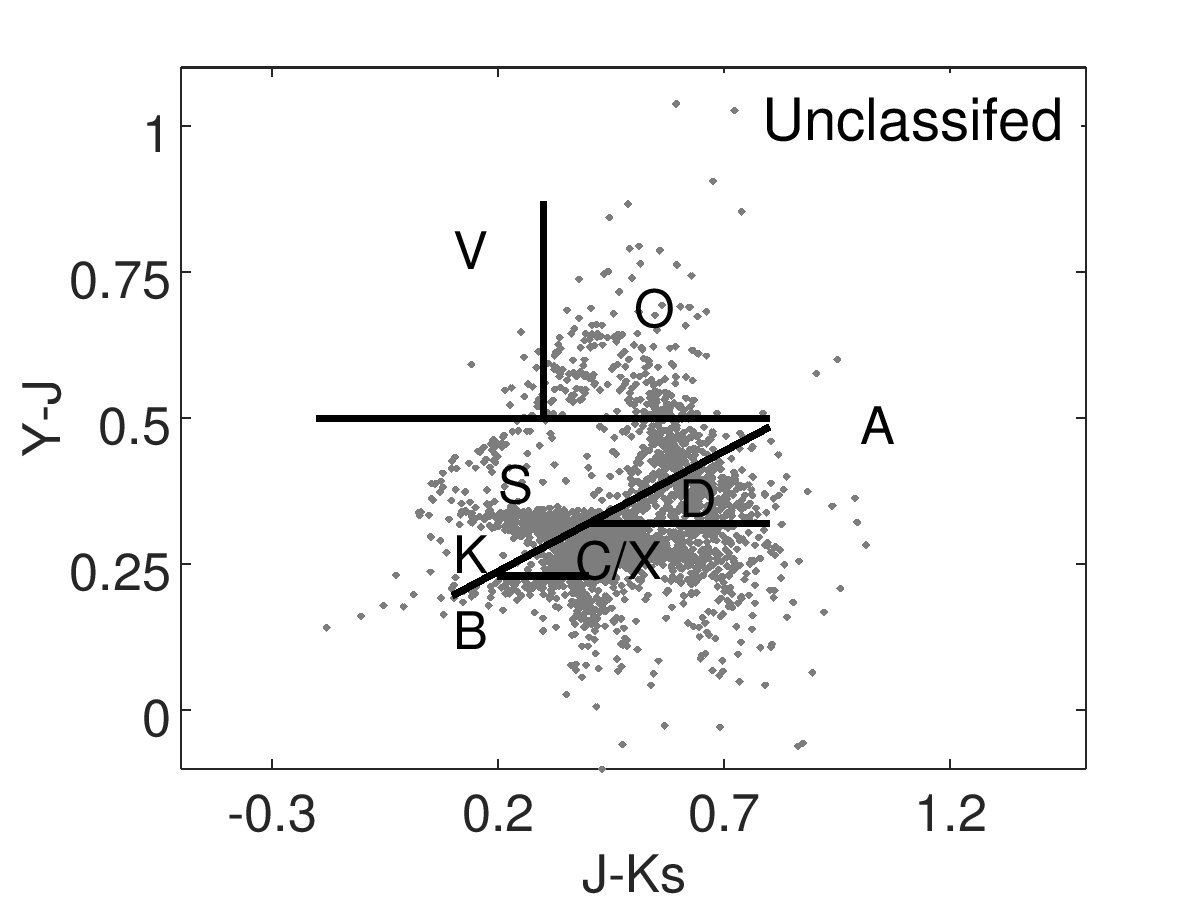}
\end{center}
\caption{ Distribution of asteroids in the (Y-J) versus (J-Ks) plot for asteroids with (Y-J)$_{err}\leq0.118$ and (J-Ks)$_{err}\leq0.136$ (asteroids with errors in Q2 limits). (Left) The objects for which a final taxonomic class was assigned are shown with blue, while those for which the two algorithms gave a different solution are shown in grey.(Right) For clarity, the unclassified asteroids are shown as a separate figure.}
\label{ClassUclass}
\end{figure*}

\section{Discussion}

The taxonomy reported here provides a large number of asteroids classified as $S^{ni}$ and $V^{ni}$. The number of objects compatible with the C- and X- complexes is only about one quarter of those compatible with S-complex. This ratio is a result of several biases introduced because of the observational limits and the methods used for classification. 

As a consequence of limit in the apparent magnitude of the VISTA-VHS survey, objects that have high albedo and are close to the observer have a higher probability of being observed \citep{2002aste.book...71J}. This introduces a bias favoring our $S^{ni}$ and $V^{ni}$ groups, as well as the inner main belt asteroids, which have the largest fraction of rocky, S-type asteroids. The observational bias can be roughly outlined by considering the number of objects above and below the $\alpha$ line \citep[see Eq.~3 from ][]{2016A&A...591A.115P} which is an approximation to separate the S- and V- from the B-, C-, X-, D- types. This implies a ratio of  $\sim60\%$ for possible detections of S- and V- types.

The second bias is introduced by the classification method. While the $S^{ni}$ and $V^{ni}$ groups are well defined in the color space, the $B_k^{ni}$, $C^{ni}$, $C_gx^{ni}$, $X_t^{ni}$, $K_l^{ni}$, and $D_s^{ni}$ groups partially overlap at certain intervals. In most of the cases where the borders are not well constrained (as is the case for these six groups), the two algorithms used for the classification provide different solutions and the objects are unclassified. These six groups were considered as a single cluster by \citet{Morate2018} and used to analyze the compositional distribution across the asteroids families. 

A subtle bias is also introduced by the reference set used for the definition of the classes. To define X-complex types, \citet{2009Icar..202..160D} used only 32 asteroids, 45 for C-complex, and 198 for S-complex. The number of objects classified by us using MOVIS catalog follows roughly the same proportion: 91 $X_t^{ni}$, 197 $C_{gx}^{ni}$, 613 $C^{ni}$, and 3\,315 $S^{ni}$.

About one third (2\,601 out of 9\,097) of the objects with errors in Q2 limits were not uniquely assigned to any of the taxonomical groups. In this cases the two algorithms provided different classifications. A fraction of this unclassified sample corresponds to those asteroids that have color values at the border of or in the overlapping interval between multiple groups. Most of these cases are in the region of the $B_k^{ni}$, $C^{ni}$, $C_gx^{ni}$, $X_t^{ni}$, and $K_l^{ni}$ groups. There is also a cluster of unclassified objects between $A_d^{ni}$ and $D_s^{ni}$ groups. The distribution of classified and unclassified asteroids is shown in Fig.~\ref{ClassUclass}. 

There is a sample of unclassified objects that were not categorized in any of the groups we defined. As an example, objects with (Y-J)$>0.5$ and intermediate (J-Ks) values, i.e., $0.3<(J-Ks)$ $\lesssim0.65$, were not classified. The most likely spectral types that match them from the reference set are the O and R types, which were not used in this study as they are defined on a single spectrum. Thus, about 218 objects fall in this category. Their distribution is widely spread in the color space.

\begin{figure}
\begin{center}
\includegraphics[width=9cm]{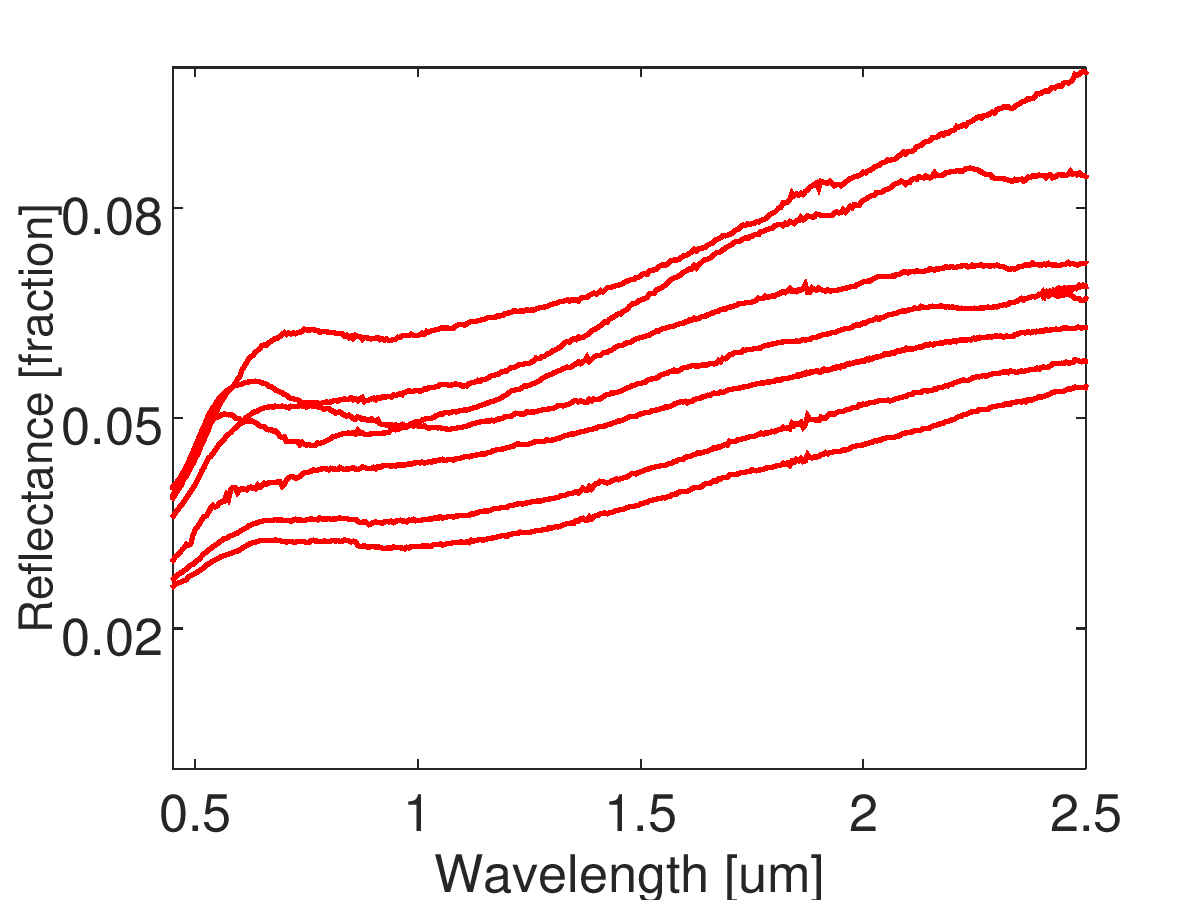}
\end{center}
\caption{ Examples of visible to near-infrared reflectance spectra of CM2 carbonaceous chondrites. The reflectances are shown in absolute units (i.e. fraction of reflected light). The spectra are almost flat in Y and J  bands and then turn extremely red in J and Ks bands (almost doubling their reflectance value). These type of compositions may explain the unclassified group with (Y-J)$\approx0.3$ and  (J-Ks)$\geq \sim0.6$.}
\label{CM2}
\end{figure}

Another set of unclassified objects (308 asteroids) lies in between the regions occupied by $X_t^{ni}$, $A_d^{ni}$, and $D_s^{ni}$. This corresponds to values of (Y-J)$~\leq\sim0.3$ and $0.55<(J-Ks)$. These colors indicate an almost flat spectral curve in the Y and J bands that turns red beyond 1.25 $\mu$m. About $\sim$65\% of the WISE albedo values measured for 207 out of these 308 unclassified objects are lower than 0.1. Meteoritic samples with equivalent spectral behavior are CM2 carbonaceous chondrites (Fig.~\ref{CM2}), and in particular samples of Murchison meteorite heated at 400 - 500 $^\circ C$. Their reflectances at 0.55 $\mu m$  are comparable with the visible albedo of the asteroids.

\begin{figure}
\begin{center}
\includegraphics[width=9cm]{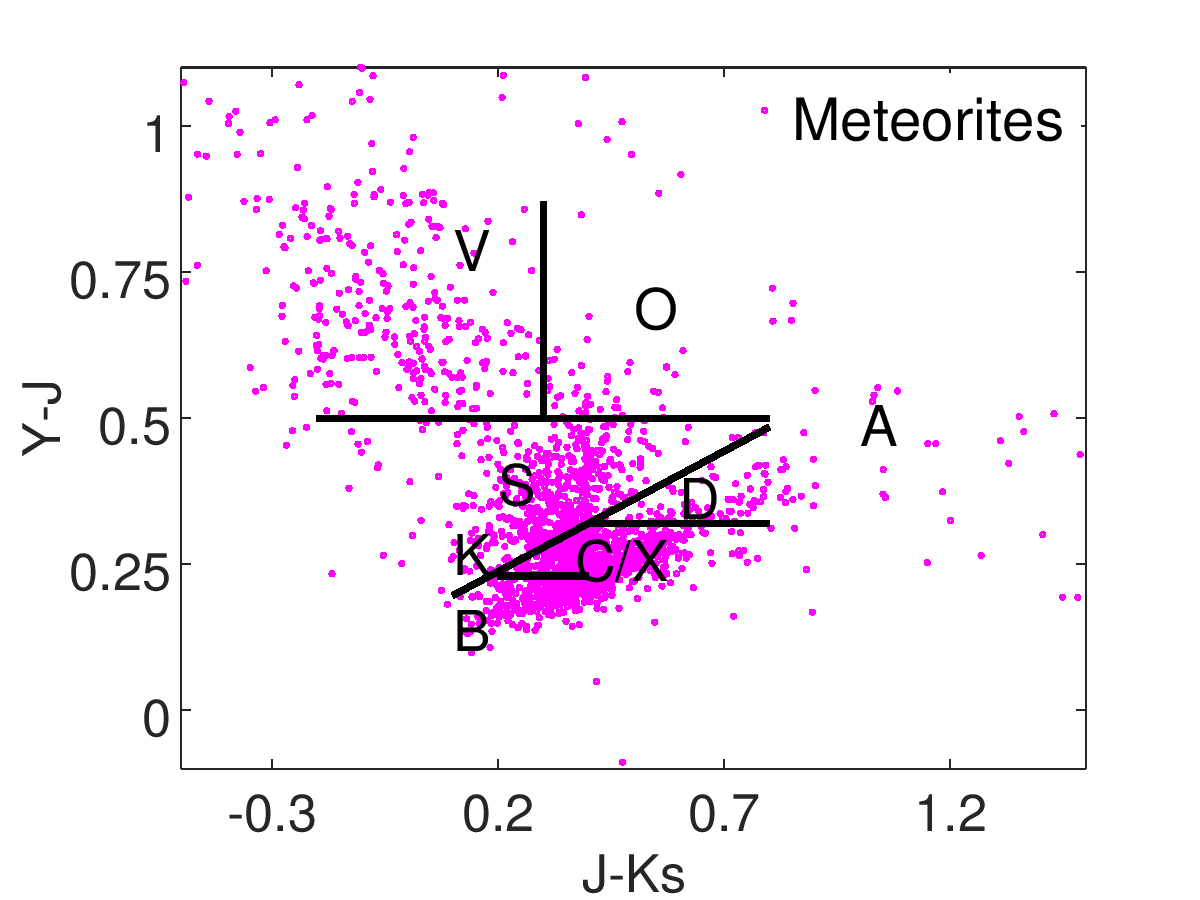}
\end{center}
\caption{NIR colors computed from spectra of all meteorites currently available in the RELAB database. A total number of 2\,636 spectra was used.}
\label{Metcolor}
\end{figure}

To assess how the existing meteorites in our collections sample the asteroids compositions we computed the synthetic (Y-J) and (J-Ks) colors for spectral data available in the RELAB database\footnote{\url{http://www.planetary.brown.edu/relab/}} \citep{2004LPI....35.1720P}. The results are shown in Fig.~\ref{Metcolor}. The asteroids distribution in the color-color space follows the same pattern as that of meteorites, but with a much broader distribution. This reflects the known fact that our meteorite records do not sample all the compositions observed in asteroids.

NIR colors are efficient for identifying end-member classes such as V-types and A-types. Also, the identified objects have albedo values in agreement with the two taxonomical classes (thus, they are less affected by observational biases). The V-type and the A-type candidates are considered fragments of differentiated bodies. Thus, they play a key role for understanding the accretion and geochemical evolution in the main belt \citep{1996M&PS...31..607B, 2015aste.book..533S, 2015aste.book..573S}. 

The V-types are associated with basaltic achondrite meteorites \citep[e.g.][]{1970Sci...168.1445M,1977GeCoA..41.1271C} which have metal-free pyroxene dominated composition and are representative for crustal material \citep{1996M&PS...31..607B}. The olivine dominated asteroids,  classified as A-types, are expected to be formed through magmatic differentiation \citep[e.g.][]{1998JGR...10313675S, 2004A&A...422L..59D, 2007M&PS...42..155S, 2014Icar..228..288S}. Therefore they could be mantle fragments of differentiated bodies \citep[e.g.][]{1996M&PS...31..607B}, or come from nebular processes which produce olivine-dominated objects similar to R-chondrites \citep{1994Metic..29..275S}. 

Our current meteoritic evidences show that at least 100 chondritic parent bodies in the main belt experienced partial or complete melting and differentiation before having been disrupted \citep[][and references there in]{1996M&PS...31..607B, 2014Icar..228..288S, 2015aste.book..573S}. Therefore, they should have produced a much larger number of olivine-dominated objects. The spectral surveys show a paucity of such bodies, which is known as the "missing mantle problem" or the "olivine paradox" \citep{2015aste.book...13D}. 

The hypotheses proposed to explain the olivine paradox vary from inferring the small size of olivine dominated asteroids (below the limit of detection), i.e., the "battered to bits" scenario \citep{1996M&PS...31..607B}, to those suggesting that the classic view of asteroids differentiating into a pyroxene-rich crust, olivine-rich mantle, and iron core may be uncommon \citep{2011E&PSL.305....1E}.

Our dataset allows to quantify this problem up to a $\sim2$ km size limit of the bodies. Both, V- and A-types, have prominent features, which put them in well defined regions in (Y-J) vs. (J-Ks) color-color space. V-types have approximately (Y-J) $\geq  0.5$ and (J-Ks) $\leq 0.3$ ($V^{ni}$). A-types are characterized by a red spectral slope which in terms of colors is equivalent to (J-Ks) $\geq \sim 0.65$ ($A_d^{ni}$). Fig.~\ref{ClassUclass} shows a significant difference in terms of population density over the two color regions, with the V-types being significantly more abundant than the A-types.

\begin{table}
\caption{ List of asteroids classified as $A_d^{ni}$ based on MOVIS-C near-infrared colors and having $p_V\geq0.15$ \citep{2011ApJ...743..156M, 2014ApJ...792...30M}.}

\begin{tabular}{l l l l l}                                                                                                                               
\hline                                                                                                                               
\hline
2872 &13734&45785&73257 &125725\\
3323 &14504&46155&75629 &125969\\
4982 &15291&47269&75709 &132542\\
6319 &17059&49075&78174 &133572\\
6323 &17302&50115&83869 &133771\\
6754 &17671&53697&84903 &139502\\
6900 &21241&54252&86166 &144419\\
7032 &26533&57501&88405 &152899\\
7225 &28858&58339&91843 &153884\\
8730 &31935&60774&95560 &163913\\
9690 &34084&60888&97880 &200449\\
9720 &36405&61972&99633 &238720\\
12259&37752&62046&121232& -    \\
12308&44265&69803&123980& -    \\
\hline                                                                                                                                                                                                               
\end{tabular}
\label{ConfirmedAtype}                                                                                                   
\end{table}

In order to add more constraints to the quality of our dataset, we consider only those objects which have measured albedo $p_V\geq0.15$. This allows to identify 68 A-type (Table~\ref{ConfirmedAtype}) and 338 V-type candidates (including 101 members of the Vesta family). This gives a ratio of about five times more basaltic asteroids than the olivine dominated objects (about four if the Vesta family members are not considered). 

\begin{figure}
\begin{center}
\includegraphics[width=9cm]{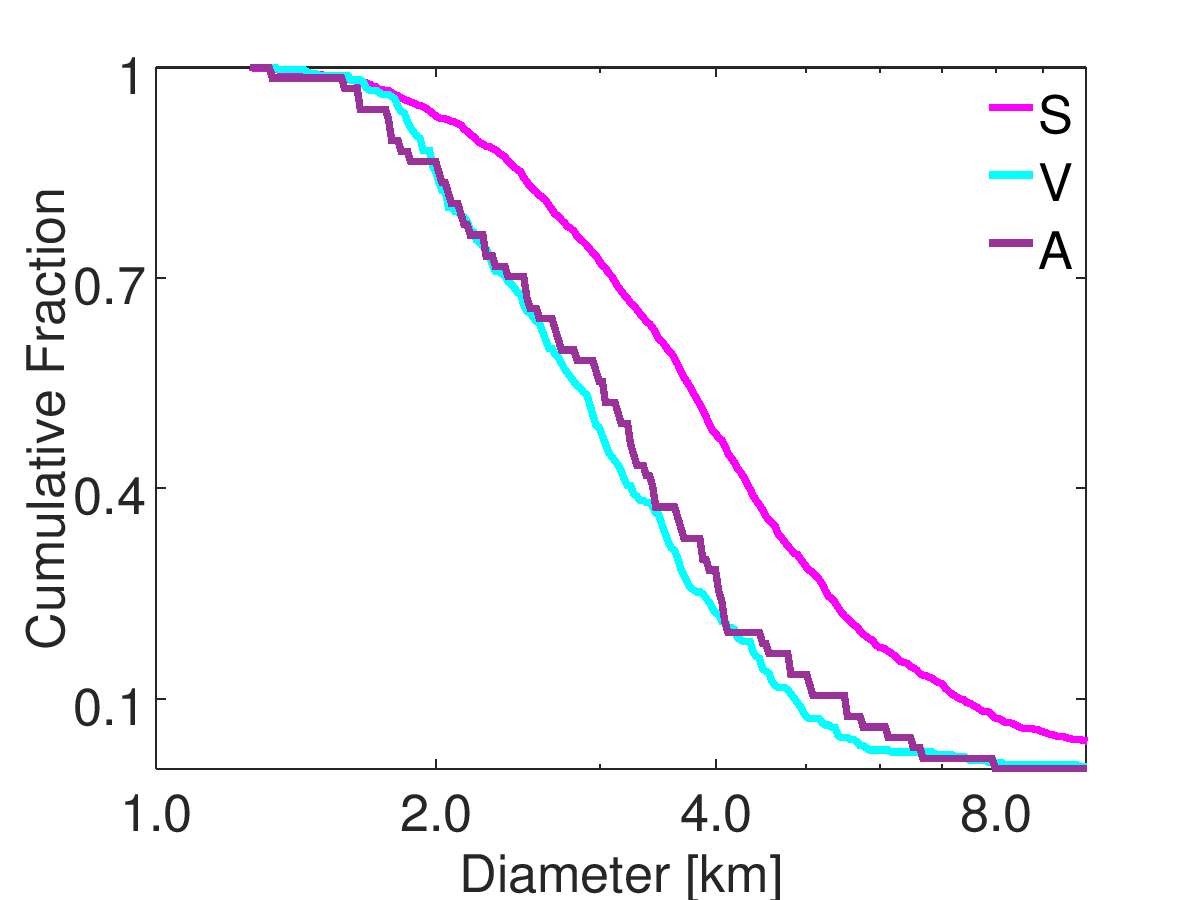}
\end{center}
\caption{Size-frequency distribution of asteroids classified as $A_d^{ni}$, $V^{ni}$, and $S^{ni}$ and with a measured albedo value $p_V\geq0.15$.}
\label{SFD}
\end{figure}

The diameters of A-type candidates range from 1.3 km for (125969) 2003 FU30 to 8 km for (4982) Bartini, with an average value of $D_{Aavg} = 3.4\pm1.4$ km. The exception is (2872) Gentelec which has a diameter of 13.6 km for a $p_V=0.15\pm0.016$, and it was classified as D-type by \citet{2002Icar..158..146B}, while we reported it as $A_d^{ni}$. A similar average diameter value is computed for V-type candidates, $D_{Vavg} = 3.5\pm4.3$ km. Also, the size-frequency distributions are identical for the two classes (Fig.~\ref{SFD}), and significantly different from the S-types (associated with ordinary chondrites) which have similar albedos. Therefore the different ratio of A-types and V-types candidates, and a similar size-frequency distribution of the two groups is not in favor of the two scenarios outlined by \citet{1996M&PS...31..607B}. 

One interesting case in our dataset is asteroid (11616) 1996 BQ2 which has orbital parameters that place it in the Cybele population. Its (Y-J) = $0.37\pm0.05$ and (J-Ks)$~= 0.79\pm0.07$ colors suggest an A-type candidate. The probabilistic algorithm report it as $A_d^{ni}$ with a probability p = 0.93 and the KNN algorithm found it as $D_s^{ni}$ with probability p = 0.58 and as $A_d^{ni}$ with p = 0.42. It is classified as an S-type by \citet{2010A&A...510A..43C}, according to its SDSS visible photometry, and its WISE albedo is $p_V = 0.246\pm0.038$. The broad observational measurements make (11616) 1996 BQ2 an olivine-dominated candidate\footnote{ We have just received words of the confirmation as A-type based on NIR spectral data by F.~E. DeMeo and collaborators (F.~E. DeMeo personal communication)}, in a region dominated by P and D types \citep{1989asteconf..316G}. The presence of this object may not be so unexpected since \citet{2003Icar..162...10M} noted that the number of S-type objects near the inner side of the Hecuba gap (i.e., the 2:1 mean motion resonance with Jupiter) is low but not zero. For example, based on SDSS data, \citet{2015MNRAS.451..244C} report five S-types, and one A-type in the Cybele region. Also, \citet{2010Icar..206..729G} reported (11616) 1996 BQ2 as an S-type and hypothesized that the presence of these objects in the Cybele population could be the result of a dynamical mechanism influencing objects in the inner border of the gap, like migration due to the Yarkovsky effect, or resonance perturbations on unstable asteroids inside the Hecuba gap.

\section{Future perspectives}

Taxonomic classifications can also be inferred from other parameters than from spectral and spectro-photometric observations. \citet{2012Icar..219..283O} searched for correlations between the classification of asteroids and photometric phase curve using the $H,~G_{12}$ photometric phase function. They found that the $G_{12}$ parameter provides a statistical approach for studying taxonomic classification for large population of asteroids. \citet{2017Icar..284...30B} showed that by using polarimetric data it is possible to refine asteroids taxonomy by distinguishing several types which can not be identified using spectra over the visible to NIR interval. Moreover, based on the similarity of the polarization phase curves for asteroids with the same taxonomic type they suggest that polarimetric behavior is intimately related to the surface composition. 

Several surveys will provide a large amount of data in the next decade. In particular, the high-cadence multi-epoch observations of the Large Synoptic Survey Telescope \citep{2009arXiv0912.0201L} will dramatically improve our knowledge of solar system bodies through discoveries and orbit determination \citep{2017ApJS..233...21R}. Besides the astrometric positions, the obtained photometry  will allow phase curves studies for compositional types \citep{2012Icar..219..283O}, and shape modeling \citep{2005EM&P...97..179D,2015aste.book..183D}. 

Spectroscopically, it is expected that a robust classification can be obtained for at least 100\,000 asteroids from the low-resolution  visible spectra obtained by Gaia \citep{2012P&SS...73...86D}.

Also, during their nominal execution times the J-PLUS/S-PLUS photometric surveys -- a joint Spanish-Brazilian project that aims to map an area of the sky of 8\,000 square degrees, are expected to observe tens of thousands of asteroids and measure their visible colors using a 12-filter system covering the 0.3--1.0 $\mu m$ range \citep{2014acm..conf..429D}.

New data in the V-NIR region are also expected from Euclid survey. This is a wide-field space mission, equipped  with a silicon-carbide 1.2 m aperture Korsch telescope and two instruments, a visible imaging camera, and a Near Infrared Spectrometer and Photometer \citep{2011arXiv1110.3193L}. By considering the current survey design, \citet{2018A&A...609A.113C} approximate that about about 150\,000 solar system objects, mainly from the asteroid main-belt, should be observable by Euclid.

The faint magnitude limit of this survey (photometrically, $V_{AB}\approx 24.5$), its spectral coverage (visible to near-infrared considering spectral and spectro-photometric data), and the survey  strategy (survey will avoid ecliptic latitudes and the photometric pattern consists in repeated sequences of four broadband filters) will open a window to peculiar classes of objects (small objects with high inclinations). In particular, as shown in this paper, the use of Y and J filters provides the means to infer the main compositional groups.

Each observation method provides a piece of the puzzle for transforming  the "star-like" appearance of the asteroids into a little world. The meaning of each observation  and the information that it brings depend on the spectral characteristics that it is sampling. In this context, taxonomy provides a common language for discussing the asteroids compositional properties. By combining the large quantity of data provided by the surveys, an unprecedented map of the minor bodies population will be revealed.

\section{Conclusions}

We have inferred a taxonomic classification for the 18\,265 asteroids catalogued by MOVIS using a probabilistic method and k-nearest neighbors algorithm. This database provides the largest NIR existing set of colors for solar system objects. They were obtained with Y, J, H, and Ks filters in the framework of VISTA-VHS survey, using the VHSv20161007 intermediate release.

The asteroids spectrally classified by \citet{2009Icar..202..160D} were considered as a reference set. The 24 classes defined by Bus-DeMeo taxonomy were clustered in several groups by considering similar compositional types and their distance in the $\{(Y-J), (J-Ks), (H-Ks)\}$ space. The taxonomic classification reported here is based on these classes.

The applied algorithms provided a taxonomic type for all objects with at least (Y-J) and (J-Ks) colors observed. The final classifications is reported for a set of 6\,494 asteroids which have (Y-J)$_{err}\leq0.118$ and (J-Ks)$_{err}\leq0.136$ and were assigned to the same taxonomical class by the KNN and probabilistic algorithms. These asteroids were distributed as follows:  144 bodies classified as $B_k^{ni}$, 613 as $C^{ni}$, 197 as $C_{gx}^{ni}$, 91 as $X_t^{ni}$, 440 as $D_s^{ni}$, 665 as $K_l^{ni}$, 233 as $A_d^{ni}$, 3\,315 as $S^{ni}$, and 798 as $V^{ni}$. Here, the superscript indicates that the classification was obtained based on the NIR colors and the subscript indicates possible miss-identifications with other types.

By comparing our classification with the one reported by \citet{2010A&A...510A..43C} based on visible colors from SDSS survey we found a $\sim70\%$ of matching for identifying silicaceous and carbonaceous types.

We also found that the asteroid population presents a broader range of NIR colors than those computed from meteorite spectra available in the RELAB database. This points towards a wider compositional diversity of asteroids not sampled by our current meteorite records.

We reported the albedo distribution for each taxonomic group based on the data existing in WISE. By adding the albedo constraint we found that our V-type and A-type candidates have identical size-frequency distributions but a ratio of one to five in favor of V-types. This result adds additional constraints for the hypothesis of "missing mantle problem".

Taxonomic classification of asteroids catalogued by MOVIS provides the opportunity to obtain large scale distribution for asteroidal population, to study the faint objects and to select targets for detailed spectral investigations. 

\begin{acknowledgements}
We want to specially thank to S.~J. Bus and F.~E. DeMeo for sharing  the 371 spectra they used to define the Bus-DeMeo taxonomy.
This research utilizes spectra acquired with the NASA RELAB facility at Brown University.   The data has been numerically analyzed with GNU Octave, TOPCAT(\url{http://www.starlink.ac.uk/topcat/}) and Python Scikit-learn.
MP acknowledges support from the AYA2015-67772-R (MINECO, Spain). The work of MP was also supported by a grant of the Romanian National Authority for Scientific Research - UEFISCDI, project number PN-III-P1-1.2-PCCDI-2017-0371.  DM, JL, and JdL acknowledge support from the project AYA2012-39115-C03-03 and ESP2013-47816-C4-2-P (MINECO). JdL acknowledges financial support from MINECO under the 2015 Severo Ochoa Program MINECO SEV-2015-0548. DM gratefully acknowledges the Spanish Ministry of Economy and Competitiveness (MINECO) for the financial support received in the form of a Severo-Ochoa PhD fellowship, within the Severo-Ochoa Interna-tional PhD Program. The work of ILB was supported by a grant of the Romanian National Authority for Scientific Research - UEFISCDI, project number PN-III-P1-1.2-PCCDI-2017-0371. { We thank Dr. Michael Mommert for his constructive and helpful suggestions.}
\end{acknowledgements}

\bibliographystyle{aa}
\bibliography{MOVISTAX}

\begin{thebibliography}{90}
\expandafter\ifx\csname natexlab\endcsname\relax\def\natexlab#1{#1}\fi

\bibitem[{{Al{\'{\i}}-Lagoa} {et~al.}(2013){Al{\'{\i}}-Lagoa}, {de Le{\'o}n},
  {Licandro}, {Delb{\'o}}, {Campins}, {Pinilla-Alonso}, \&
  {Kelley}}]{Ali-Lagoa:2013aa}
{Al{\'{\i}}-Lagoa}, V., {de Le{\'o}n}, J., {Licandro}, J., {et~al.} 2013, \aap,
  554, A71

\bibitem[{{Baudrand} {et~al.}(2001){Baudrand}, {Bec-Borsenberger},
  {Borsenberger}, \& {Barucci}}]{2001A&A...375..275B}
{Baudrand}, A., {Bec-Borsenberger}, A., {Borsenberger}, J., \& {Barucci}, M.~A.
  2001, \aap, 375, 275

\bibitem[{{Belskaya} {et~al.}(2017){Belskaya}, {Fornasier}, {Tozzi},
  {Gil-Hutton}, {Cellino}, {Antonyuk}, {Krugly}, {Dovgopol}, \&
  {Faggi}}]{2017Icar..284...30B}
{Belskaya}, I.~N., {Fornasier}, S., {Tozzi}, G.~P., {et~al.} 2017, \icarus,
  284, 30

\bibitem[{{Bottke} {et~al.}(2002){Bottke}, {Vokrouhlick{\'y}}, {Rubincam}, \&
  {Broz}}]{2002aste.book..395B}
{Bottke}, Jr., W.~F., {Vokrouhlick{\'y}}, D., {Rubincam}, D.~P., \& {Broz}, M.
  2002, {The Effect of Yarkovsky Thermal Forces on the Dynamical Evolution of
  Asteroids and Meteoroids}, ed. W.~F. {Bottke}, Jr., A.~{Cellino},
  P.~{Paolicchi}, \& R.~P. {Binzel}, 395--408

\bibitem[{{Bottke} {et~al.}(2006){Bottke}, {Vokrouhlick{\'y}}, {Rubincam}, \&
  {Nesvorn{\'y}}}]{Bottke:2006aa}
{Bottke}, Jr., W.~F., {Vokrouhlick{\'y}}, D., {Rubincam}, D.~P., \&
  {Nesvorn{\'y}}, D. 2006, Annual Review of Earth and Planetary Sciences, 34,
  157

\bibitem[{{Bowell} {et~al.}(1978){Bowell}, {Chapman}, {Gradie}, {Morrison}, \&
  {Zellner}}]{1978Icar...35..313B}
{Bowell}, E., {Chapman}, C.~R., {Gradie}, J.~C., {Morrison}, D., \& {Zellner},
  B. 1978, \icarus, 35, 313

\bibitem[{{Brasil} {et~al.}(2017){Brasil}, {Roig}, {Nesvorn{\'y}}, \&
  {Carruba}}]{2017MNRAS.468.1236B}
{Brasil}, P.~I.~O., {Roig}, F., {Nesvorn{\'y}}, D., \& {Carruba}, V. 2017,
  \mnras, 468, 1236

\bibitem[{Buitinck {et~al.}(2013)Buitinck, Louppe, Blondel, Pedregosa, Mueller,
  Grisel, Niculae, Prettenhofer, Gramfort, Grobler, Layton, VanderPlas, Joly,
  Holt, \& Varoquaux}]{sklearnapi}
Buitinck, L., Louppe, G., Blondel, M., {et~al.} 2013, in ECML PKDD Workshop:
  Languages for Data Mining and Machine Learning, 108--122

\bibitem[{{Burbine} {et~al.}(1996){Burbine}, {Meibom}, \&
  {Binzel}}]{1996M&PS...31..607B}
{Burbine}, T.~H., {Meibom}, A., \& {Binzel}, R.~P. 1996, Meteoritics and
  Planetary Science, 31, 607

\bibitem[{{Bus} \& {Binzel}(2002{\natexlab{a}})}]{2002Icar..158..146B}
{Bus}, S.~J. \& {Binzel}, R.~P. 2002{\natexlab{a}}, \icarus, 158, 146

\bibitem[{{Bus} \& {Binzel}(2002{\natexlab{b}})}]{2002Icar..158..106B}
{Bus}, S.~J. \& {Binzel}, R.~P. 2002{\natexlab{b}}, \icarus, 158, 106

\bibitem[{{Campins} {et~al.}(2010){Campins}, {Morbidelli}, {Tsiganis}, {de
  Le{\'o}n}, {Licandro}, \& {Lauretta}}]{2010ApJ...721L..53C}
{Campins}, H., {Morbidelli}, A., {Tsiganis}, K., {et~al.} 2010, \apjl, 721, L53

\bibitem[{{Carruba} {et~al.}(2014){Carruba}, {Huaman}, {Domingos}, {Santos}, \&
  {Souami}}]{2014MNRAS.439.3168C}
{Carruba}, V., {Huaman}, M.~E., {Domingos}, R.~C., {Santos}, C.~R.~D., \&
  {Souami}, D. 2014, \mnras, 439, 3168

\bibitem[{{Carruba} {et~al.}(2015){Carruba}, {Nesvorn{\'y}}, {Aljbaae}, \&
  {Huaman}}]{2015MNRAS.451..244C}
{Carruba}, V., {Nesvorn{\'y}}, D., {Aljbaae}, S., \& {Huaman}, M.~E. 2015,
  \mnras, 451, 244

\bibitem[{{Carry}(2018)}]{2018A&A...609A.113C}
{Carry}, B. 2018, \aap, 609, A113

\bibitem[{{Carry} {et~al.}(2016){Carry}, {Solano}, {Eggl}, \&
  {DeMeo}}]{2016Icar..268..340C}
{Carry}, B., {Solano}, E., {Eggl}, S., \& {DeMeo}, F.~E. 2016, \icarus, 268,
  340

\bibitem[{{Carvano} {et~al.}(2010){Carvano}, {Hasselmann}, {Lazzaro}, \&
  {Moth{\'e}-Diniz}}]{2010A&A...510A..43C}
{Carvano}, J.~M., {Hasselmann}, P.~H., {Lazzaro}, D., \& {Moth{\'e}-Diniz}, T.
  2010, \aap, 510, A43

\bibitem[{{Casagrande} {et~al.}(2012){Casagrande}, {Ram{\'{\i}}rez},
  {Mel{\'e}ndez}, \& {Asplund}}]{2012ApJ...761...16C}
{Casagrande}, L., {Ram{\'{\i}}rez}, I., {Mel{\'e}ndez}, J., \& {Asplund}, M.
  2012, \apj, 761, 16

\bibitem[{{Chapman} {et~al.}(1975){Chapman}, {Morrison}, \&
  {Zellner}}]{1975Icar...25..104C}
{Chapman}, C.~R., {Morrison}, D., \& {Zellner}, B. 1975, \icarus, 25, 104

\bibitem[{{Clark} {et~al.}(2010){Clark}, {Ziffer}, {Nesvorny}, {Campins},
  {Rivkin}, {Hiroi}, {Barucci}, {Fulchignoni}, {Binzel}, {Fornasier}, {DeMeo},
  {Ockert-Bell}, {Licandro}, \& {Moth{\'e}-Diniz}}]{2010JGRE..115.6005C}
{Clark}, B.~E., {Ziffer}, J., {Nesvorny}, D., {et~al.} 2010, Journal of
  Geophysical Research (Planets), 115, E06005

\bibitem[{{Cloutis} {et~al.}(2010){Cloutis}, {Hardersen}, {Bish}, {Bailey},
  {Gaffey}, \& {Craig}}]{2010M&PS...45..304C}
{Cloutis}, E.~A., {Hardersen}, P.~S., {Bish}, D.~L., {et~al.} 2010, Meteoritics
  and Planetary Science, 45, 304

\bibitem[{{Consolmagno} \& {Drake}(1977)}]{1977GeCoA..41.1271C}
{Consolmagno}, G.~J. \& {Drake}, M.~J. 1977, \gca, 41, 1271

\bibitem[{{Cross} {et~al.}(2012){Cross}, {Collins}, {Mann}, {Read}, {Sutorius},
  {Blake}, {Holliman}, {Hambly}, {Emerson}, {Lawrence}, \&
  {Noddle}}]{2012A&A...548A.119C}
{Cross}, N.~J.~G., {Collins}, R.~S., {Mann}, R.~G., {et~al.} 2012, \aap, 548,
  A119

\bibitem[{{de Le{\'o}n} {et~al.}(2004){de Le{\'o}n}, {Duffard}, {Licandro}, \&
  {Lazzaro}}]{2004A&A...422L..59D}
{de Le{\'o}n}, J., {Duffard}, R., {Licandro}, J., \& {Lazzaro}, D. 2004, \aap,
  422, L59

\bibitem[{{de Le{\'o}n} {et~al.}(2012){de Le{\'o}n}, {Pinilla-Alonso},
  {Campins}, {Licandro}, \& {Marzo}}]{2012Icar..218..196D}
{de Le{\'o}n}, J., {Pinilla-Alonso}, N., {Campins}, H., {Licandro}, J., \&
  {Marzo}, G.~A. 2012, \icarus, 218, 196

\bibitem[{{De Pr{\'a}} \& {Carvano}(2014)}]{2014acm..conf..429D}
{De Pr{\'a}}, M. \& {Carvano}, J. 2014, in Asteroids, Comets, Meteors 2014, ed.
  K.~{Muinonen}, A.~{Penttil{\"a}}, M.~{Granvik}, A.~{Virkki}, G.~{Fedorets},
  O.~{Wilkman}, \& T.~{Kohout}

\bibitem[{{Delbo'} {et~al.}(2012){Delbo'}, {Gayon-Markt}, {Busso}, {Brown},
  {Galluccio}, {Ordenovic}, {Bendjoya}, \& {Tanga}}]{2012P&SS...73...86D}
{Delbo'}, M., {Gayon-Markt}, J., {Busso}, G., {et~al.} 2012, \planss, 73, 86

\bibitem[{{DeMeo} {et~al.}(2015){DeMeo}, {Alexander}, {Walsh}, {Chapman}, \&
  {Binzel}}]{2015aste.book...13D}
{DeMeo}, F.~E., {Alexander}, C.~M.~O., {Walsh}, K.~J., {Chapman}, C.~R., \&
  {Binzel}, R.~P. 2015, in Asteroids IV, ed. P.~{Michel}, F.~E. {DeMeo}, \&
  W.~F. {Bottke}, 13--41

\bibitem[{{DeMeo} {et~al.}(2014){DeMeo}, {Binzel}, {Carry}, {Polishook}, \&
  {Moskovitz}}]{2014Icar..229..392D}
{DeMeo}, F.~E., {Binzel}, R.~P., {Carry}, B., {Polishook}, D., \& {Moskovitz},
  N.~A. 2014, \icarus, 229, 392

\bibitem[{{DeMeo} {et~al.}(2009){DeMeo}, {Binzel}, {Slivan}, \&
  {Bus}}]{2009Icar..202..160D}
{DeMeo}, F.~E., {Binzel}, R.~P., {Slivan}, S.~M., \& {Bus}, S.~J. 2009,
  \icarus, 202, 160

\bibitem[{{DeMeo} \& {Carry}(2013)}]{2013Icar..226..723D}
{DeMeo}, F.~E. \& {Carry}, B. 2013, \icarus, 226, 723

\bibitem[{{DeMeo} \& {Carry}(2014)}]{2014Natur.505..629D}
{DeMeo}, F.~E. \& {Carry}, B. 2014, \nat, 505, 629

\bibitem[{{Durech} {et~al.}(2015){Durech}, {Carry}, {Delbo}, {Kaasalainen}, \&
  {Viikinkoski}}]{2015aste.book..183D}
{Durech}, J., {Carry}, B., {Delbo}, M., {Kaasalainen}, M., \& {Viikinkoski}, M.
  2015, in Asteroids IV, ed. P.~{Michel}, F.~E. {DeMeo}, \& W.~F. {Bottke},
  183--202

\bibitem[{{Elkins-Tanton} {et~al.}(2011){Elkins-Tanton}, {Weiss}, \&
  {Zuber}}]{2011E&PSL.305....1E}
{Elkins-Tanton}, L.~T., {Weiss}, B.~P., \& {Zuber}, M.~T. 2011, Earth and
  Planetary Science Letters, 305, 1

\bibitem[{{Emerson} {et~al.}(2004){Emerson}, {Irwin}, {Lewis}, {Hodgkin},
  {Evans}, {Bunclark}, {McMahon}, {Hambly}, {Mann}, {Bond}, {Sutorius}, {Read},
  {Williams}, {Lawrence}, \& {Stewart}}]{2004SPIE.5493..401E}
{Emerson}, J.~P., {Irwin}, M.~J., {Lewis}, J., {et~al.} 2004, in Society of
  Photo-Optical Instrumentation Engineers (SPIE) Conference Series, Vol. 5493,
  Optimizing Scientific Return for Astronomy through Information Technologies,
  ed. P.~J. {Quinn} \& A.~{Bridger}, 401--410

\bibitem[{{Fornasier} {et~al.}(2010){Fornasier}, {Clark}, {Dotto},
  {Migliorini}, {Ockert-Bell}, \& {Barucci}}]{2010Icar..210..655F}
{Fornasier}, S., {Clark}, B.~E., {Dotto}, E., {et~al.} 2010, \icarus, 210, 655

\bibitem[{{Gil-Hutton} \& {Licandro}(2010)}]{2010Icar..206..729G}
{Gil-Hutton}, R. \& {Licandro}, J. 2010, \icarus, 206, 729

\bibitem[{{Gradie} {et~al.}(1989){Gradie}, {Chapman}, \&
  {Tedesco}}]{1989asteconf..316G}
{Gradie}, J.~C., {Chapman}, C.~R., \& {Tedesco}, E.~F. 1989, in Asteroids II,
  316--335

\bibitem[{{Hahn} \& {Lagerkvist}(1988)}]{1988Icar...74..454H}
{Hahn}, G. \& {Lagerkvist}, C.-I. 1988, \icarus, 74, 454

\bibitem[{{Hambly} {et~al.}(2004){Hambly}, {Mann}, {Bond}, {Sutorius}, {Read},
  {Williams}, {Lawrence}, \& {Emerson}}]{2004SPIE.5493..423H}
{Hambly}, N.~C., {Mann}, R.~G., {Bond}, I., {et~al.} 2004, in Society of
  Photo-Optical Instrumentation Engineers (SPIE) Conference Series, Vol. 5493,
  Optimizing Scientific Return for Astronomy through Information Technologies,
  ed. P.~J. {Quinn} \& A.~{Bridger}, 423--431

\bibitem[{{Hasselmann} {et~al.}(2015){Hasselmann}, {Fulchignoni}, {Carvano},
  {Lazzaro}, \& {Barucci}}]{2015A&A...577A.147H}
{Hasselmann}, P.~H., {Fulchignoni}, M., {Carvano}, J.~M., {Lazzaro}, D., \&
  {Barucci}, M.~A. 2015, \aap, 577, A147

\bibitem[{{Ieva} {et~al.}(2016){Ieva}, {Dotto}, {Lazzaro}, {Perna}, {Fulvio},
  \& {Fulchignoni}}]{2016MNRAS.455.2871I}
{Ieva}, S., {Dotto}, E., {Lazzaro}, D., {et~al.} 2016, \mnras, 455, 2871

\bibitem[{{Irwin} {et~al.}(2004){Irwin}, {Lewis}, {Hodgkin}, {Bunclark},
  {Evans}, {McMahon}, {Emerson}, {Stewart}, \& {Beard}}]{2004SPIE.5493..411I}
{Irwin}, M.~J., {Lewis}, J., {Hodgkin}, S., {et~al.} 2004, in Society of
  Photo-Optical Instrumentation Engineers (SPIE) Conference Series, Vol. 5493,
  Optimizing Scientific Return for Astronomy through Information Technologies,
  ed. P.~J. {Quinn} \& A.~{Bridger}, 411--422

\bibitem[{{Ivezi{\'c}} {et~al.}(2001){Ivezi{\'c}}, {Tabachnik}, {Rafikov},
  {Lupton}, {Quinn}, {Hammergren}, {Eyer}, {Chu}, {Armstrong}, {Fan},
  {Finlator}, {Geballe}, {Gunn}, {Hennessy}, {Knapp}, {Leggett}, {Munn},
  {Pier}, {Rockosi}, {Schneider}, {Strauss}, {Yanny}, {Brinkmann}, {Csabai},
  {Hindsley}, {Kent}, {Lamb}, {Margon}, {McKay}, {Smith}, {Waddel}, {York}, \&
  {SDSS Collaboration}}]{2001AJ....122.2749I}
{Ivezi{\'c}}, {\v Z}., {Tabachnik}, S., {Rafikov}, R., {et~al.} 2001, \aj, 122,
  2749

\bibitem[{{Jedicke} {et~al.}(2002){Jedicke}, {Larsen}, \&
  {Spahr}}]{2002aste.book...71J}
{Jedicke}, R., {Larsen}, J., \& {Spahr}, T. 2002, in Asteroids III, ed. W.~F.
  {Bottke}, Jr., A.~{Cellino}, P.~{Paolicchi}, \& R.~P. {Binzel}, 71--87

\bibitem[{{Johnson} {et~al.}(1975){Johnson}, {Veeder}, {Loer}, \&
  {Matson}}]{1975ApJ...197..527J}
{Johnson}, T.~V., {Veeder}, G.~J., {Loer}, S.~J., \& {Matson}, D.~L. 1975,
  \apj, 197, 527

\bibitem[{{Lagerkvist} {et~al.}(2005){Lagerkvist}, {Moroz}, {Nathues},
  {Erikson}, {Lahulla}, {Karlsson}, \& {Dahlgren}}]{2005A&A...432..349L}
{Lagerkvist}, C.-I., {Moroz}, L., {Nathues}, A., {et~al.} 2005, \aap, 432, 349

\bibitem[{{Laureijs} {et~al.}(2011){Laureijs}, {Amiaux}, {Arduini},
  {Augu{\`e}res}, {Brinchmann}, {Cole}, {Cropper}, {Dabin}, {Duvet}, {Ealet},
  \& et~al.}]{2011arXiv1110.3193L}
{Laureijs}, R., {Amiaux}, J., {Arduini}, S., {et~al.} 2011, ArXiv e-prints
  [\eprint[arXiv]{1110.3193}]

\bibitem[{{Lauretta} {et~al.}(2017){Lauretta}, {Balram-Knutson}, {Beshore},
  {Boynton}, {Drouet d'Aubigny}, {DellaGiustina}, {Enos}, {Golish},
  {Hergenrother}, {Howell}, {Bennett}, {Morton}, {Nolan}, {Rizk}, {Roper},
  {Bartels}, {Bos}, {Dworkin}, {Highsmith}, {Lorenz}, {Lim}, {Mink}, {Moreau},
  {Nuth}, {Reuter}, {Simon}, {Bierhaus}, {Bryan}, {Ballouz}, {Barnouin},
  {Binzel}, {Bottke}, {Hamilton}, {Walsh}, {Chesley}, {Christensen}, {Clark},
  {Connolly}, {Crombie}, {Daly}, {Emery}, {McCoy}, {McMahon}, {Scheeres},
  {Messenger}, {Nakamura-Messenger}, {Righter}, \&
  {Sandford}}]{Lauretta:2017aa}
{Lauretta}, D.~S., {Balram-Knutson}, S.~S., {Beshore}, E., {et~al.} 2017, \ssr,
  212, 925

\bibitem[{{Lazzaro} {et~al.}(2004){Lazzaro}, {Angeli}, {Carvano},
  {Moth{\'e}-Diniz}, {Duffard}, \& {Florczak}}]{2004Icar..172..179L}
{Lazzaro}, D., {Angeli}, C.~A., {Carvano}, J.~M., {et~al.} 2004, \icarus, 172,
  179

\bibitem[{{Levison} {et~al.}(2009){Levison}, {Bottke}, {Gounelle},
  {Morbidelli}, {Nesvorn{\'y}}, \& {Tsiganis}}]{2009Natur.460..364L}
{Levison}, H.~F., {Bottke}, W.~F., {Gounelle}, M., {et~al.} 2009, \nat, 460,
  364

\bibitem[{{Lewis} {et~al.}(2010){Lewis}, {Irwin}, \&
  {Bunclark}}]{2010ASPC..434...91L}
{Lewis}, J.~R., {Irwin}, M., \& {Bunclark}, P. 2010, in Astronomical Society of
  the Pacific Conference Series, Vol. 434, Astronomical Data Analysis Software
  and Systems XIX, ed. Y.~{Mizumoto}, K.-I. {Morita}, \& M.~{Ohishi}, 91

\bibitem[{{Licandro} {et~al.}(2017){Licandro}, {Popescu}, {Morate}, \& {de
  Le{\'o}n}}]{2017A&A...600A.126L}
{Licandro}, J., {Popescu}, M., {Morate}, D., \& {de Le{\'o}n}, J. 2017, \aap,
  600, A126

\bibitem[{{LSST Science Collaboration} {et~al.}(2009){LSST Science
  Collaboration}, {Abell}, {Allison}, {Anderson}, {Andrew}, {Angel}, {Armus},
  {Arnett}, {Asztalos}, {Axelrod}, \& et~al.}]{2009arXiv0912.0201L}
{LSST Science Collaboration}, {Abell}, P.~A., {Allison}, J., {et~al.} 2009,
  ArXiv e-prints [\eprint[arXiv]{0912.0201}]

\bibitem[{{Mainzer} {et~al.}(2014){Mainzer}, {Bauer}, {Cutri}, {Grav},
  {Masiero}, {Beck}, {Clarkson}, {Conrow}, {Dailey}, {Eisenhardt}, {Fabinsky},
  {Fajardo-Acosta}, {Fowler}, {Gelino}, {Grillmair}, {Heinrichsen}, {Kendall},
  {Kirkpatrick}, {Liu}, {Masci}, {McCallon}, {Nugent}, {Papin}, {Rice},
  {Royer}, {Ryan}, {Sevilla}, {Sonnett}, {Stevenson}, {Thompson}, {Wheelock},
  {Wiemer}, {Wittman}, {Wright}, \& {Yan}}]{2014ApJ...792...30M}
{Mainzer}, A., {Bauer}, J., {Cutri}, R.~M., {et~al.} 2014, \apj, 792, 30

\bibitem[{{Mainzer} {et~al.}(2011{\natexlab{a}}){Mainzer}, {Grav}, {Bauer},
  {Masiero}, {McMillan}, {Cutri}, {Walker}, {Wright}, {Eisenhardt}, {Tholen},
  {Spahr}, {Jedicke}, {Denneau}, {DeBaun}, {Elsbury}, {Gautier}, {Gomillion},
  {Hand}, {Mo}, {Watkins}, {Wilkins}, {Bryngelson}, {Del Pino Molina}, {Desai},
  {G{\'o}mez Camus}, {Hidalgo}, {Konstantopoulos}, {Larsen}, {Maleszewski},
  {Malkan}, {Mauduit}, {Mullan}, {Olszewski}, {Pforr}, {Saro}, {Scotti}, \&
  {Wasserman}}]{2011ApJ...743..156M}
{Mainzer}, A., {Grav}, T., {Bauer}, J., {et~al.} 2011{\natexlab{a}}, \apj, 743,
  156

\bibitem[{{Mainzer} {et~al.}(2011{\natexlab{b}}){Mainzer}, {Grav}, {Masiero},
  {Hand}, {Bauer}, {Tholen}, {McMillan}, {Spahr}, {Cutri}, {Wright}, {Watkins},
  {Mo}, \& {Maleszewski}}]{2011ApJ...741...90M}
{Mainzer}, A., {Grav}, T., {Masiero}, J., {et~al.} 2011{\natexlab{b}}, \apj,
  741, 90

\bibitem[{{Masiero} {et~al.}(2015){Masiero}, {DeMeo}, {Kasuga}, \&
  {Parker}}]{Masiero:2015aa}
{Masiero}, J.~R., {DeMeo}, F.~E., {Kasuga}, T., \& {Parker}, A.~H. 2015, in
  Asteroids IV, ed. P.~{Michel}, F.~E. {DeMeo}, \& W.~F. {Bottke}, 323--340

\bibitem[{{Masiero} {et~al.}(2011){Masiero}, {Mainzer}, {Grav}, {Bauer},
  {Cutri}, {Dailey}, {Eisenhardt}, {McMillan}, {Spahr}, {Skrutskie}, {Tholen},
  {Walker}, {Wright}, {DeBaun}, {Elsbury}, {Gautier}, {Gomillion}, \&
  {Wilkins}}]{2011ApJ...741...68M}
{Masiero}, J.~R., {Mainzer}, A.~K., {Grav}, T., {et~al.} 2011, \apj, 741, 68

\bibitem[{{McCord} {et~al.}(1970){McCord}, {Adams}, \&
  {Johnson}}]{1970Sci...168.1445M}
{McCord}, T.~B., {Adams}, J.~B., \& {Johnson}, T.~V. 1970, Science, 168, 1445

\bibitem[{{McMahon} {et~al.}(2013){McMahon}, {Banerji}, {Gonzalez}, {Koposov},
  {Bejar}, {Lodieu}, {Rebolo}, \& {VHS Collaboration}}]{2013Msngr.154...35M}
{McMahon}, R.~G., {Banerji}, M., {Gonzalez}, E., {et~al.} 2013, The Messenger,
  154, 35

\bibitem[{{Migliorini} {et~al.}(2017){Migliorini}, {De Sanctis}, {Lazzaro}, \&
  {Ammannito}}]{2017MNRAS.464.1718M}
{Migliorini}, A., {De Sanctis}, M.~C., {Lazzaro}, D., \& {Ammannito}, E. 2017,
  \mnras, 464, 1718

\bibitem[{{Migliorini} {et~al.}(2018){Migliorini}, {De Sanctis}, {Lazzaro}, \&
  {Ammannito}}]{2018MNRAS.475..353M}
{Migliorini}, A., {De Sanctis}, M.~C., {Lazzaro}, D., \& {Ammannito}, E. 2018,
  \mnras, 475, 353

\bibitem[{{Misra} \& {Bus}(2008)}]{2008DPS....40.6003M}
{Misra}, A. \& {Bus}, S.~J. 2008, in Bulletin of the American Astronomical
  Society, Vol.~40, AAS/Division for Planetary Sciences Meeting Abstracts \#40,
  508

\bibitem[{{Mommert} {et~al.}(2016){Mommert}, {Trilling}, {Borth}, {Jedicke},
  {Butler}, {Reyes-Ruiz}, {Pichardo}, {Petersen}, {Axelrod}, \&
  {Moskovitz}}]{2016AJ....151...98M}
{Mommert}, M., {Trilling}, D.~E., {Borth}, D., {et~al.} 2016, \aj, 151, 98

\bibitem[{{Morate} {et~al.}(2018){Morate}, {Licandro}, {Popescu}, \& {de
  Le\'{o}n}}]{Morate2018}
{Morate}, D., {Licandro}, J., {Popescu}, M., \& {de Le\'{o}n}, J. 2018, \aap

\bibitem[{{Moth{\'e}-Diniz} {et~al.}(2008){Moth{\'e}-Diniz}, {Carvano}, {Bus},
  {Duffard}, \& {Burbine}}]{2008Icar..195..277M}
{Moth{\'e}-Diniz}, T., {Carvano}, J.~M., {Bus}, S.~J., {Duffard}, R., \&
  {Burbine}, T.~H. 2008, \icarus, 195, 277

\bibitem[{{Moth{\'e}-Diniz} {et~al.}(2003){Moth{\'e}-Diniz}, {Carvano}, \&
  {Lazzaro}}]{2003Icar..162...10M}
{Moth{\'e}-Diniz}, T., {Carvano}, J.~M.~{\'a}., \& {Lazzaro}, D. 2003, \icarus,
  162, 10

\bibitem[{{Nesvorn{\'y}} {et~al.}(2015){Nesvorn{\'y}}, {Bro{\v z}}, \&
  {Carruba}}]{2015aste.book..297N}
{Nesvorn{\'y}}, D., {Bro{\v z}}, M., \& {Carruba}, V. 2015, in Asteroids IV,
  ed. P.~{Michel}, F.~E. {DeMeo}, \& W.~F. {Bottke}, 297--321

\bibitem[{{Oszkiewicz} {et~al.}(2012){Oszkiewicz}, {Bowell}, {Wasserman},
  {Muinonen}, {Penttil{\"a}}, {Pieniluoma}, {Trilling}, \&
  {Thomas}}]{2012Icar..219..283O}
{Oszkiewicz}, D.~A., {Bowell}, E., {Wasserman}, L.~H., {et~al.} 2012, \icarus,
  219, 283

\bibitem[{Pedregosa {et~al.}(2011)Pedregosa, Varoquaux, Gramfort, Michel,
  Thirion, Grisel, Blondel, Prettenhofer, Weiss, Dubourg, Vanderplas, Passos,
  Cournapeau, Brucher, Perrot, \& Duchesnay}]{scikitlearn}
Pedregosa, F., Varoquaux, G., Gramfort, A., {et~al.} 2011, Journal of Machine
  Learning Research, 12, 2825

\bibitem[{{Perna} {et~al.}(2010){Perna}, {Barucci}, {Fornasier}, {DeMeo},
  {Alvarez-Candal}, {Merlin}, {Dotto}, {Doressoundiram}, \& {de
  Bergh}}]{2010A&A...510A..53P}
{Perna}, D., {Barucci}, M.~A., {Fornasier}, S., {et~al.} 2010, \aap, 510, A53

\bibitem[{{Pieters} \& {Hiroi}(2004)}]{2004LPI....35.1720P}
{Pieters}, C.~M. \& {Hiroi}, T. 2004, in Lunar and Planetary Science
  Conference, Vol.~35, Lunar and Planetary Science Conference, ed.
  S.~{Mackwell} \& E.~{Stansbery}

\bibitem[{{Popescu}(2012)}]{Popescu2012}
{Popescu}, M. 2012, Ph. D. thesis

\bibitem[{{Popescu} {et~al.}(2016){Popescu}, {Licandro}, {Morate}, {de
  Le{\'o}n}, {Nedelcu}, {Rebolo}, {McMahon}, {Gonzalez-Solares}, \&
  {Irwin}}]{2016A&A...591A.115P}
{Popescu}, M., {Licandro}, J., {Morate}, D., {et~al.} 2016, \aap, 591, A115

\bibitem[{{Rhodes} {et~al.}(2017){Rhodes}, {Nichol}, {Aubourg}, {Bean},
  {Boutigny}, {Bremer}, {Capak}, {Cardone}, {Carry}, {Conselice}, {Connolly},
  {Cuillandre}, {Hatch}, {Helou}, {Hemmati}, {Hildebrandt}, {Hlo{\v z}ek},
  {Jones}, {Kahn}, {Kiessling}, {Kitching}, {Lupton}, {Mandelbaum}, {Markovic},
  {Marshall}, {Massey}, {Maughan}, {Melchior}, {Mellier}, {Newman},
  {Robertson}, {Sauvage}, {Schrabback}, {Smith}, {Strauss}, {Taylor}, \& {Von
  Der Linden}}]{2017ApJS..233...21R}
{Rhodes}, J., {Nichol}, R.~C., {Aubourg}, {\'E}., {et~al.} 2017, \apjs, 233, 21

\bibitem[{{Roh} {et~al.}(2016){Roh}, {Moon}, {Shin}, {Lee}, \&
  {Kim}}]{2016DPS....4832526R}
{Roh}, D.-G., {Moon}, H.-K., {Shin}, M.-S., {Lee}, H.-J., \& {Kim}, M.-J. 2016,
  in AAS/Division for Planetary Sciences Meeting Abstracts, Vol.~48,
  AAS/Division for Planetary Sciences Meeting Abstracts, 325.26

\bibitem[{{Sanchez} {et~al.}(2014){Sanchez}, {Reddy}, {Kelley}, {Cloutis},
  {Bottke}, {Nesvorn{\'y}}, {Lucas}, {Hardersen}, {Gaffey}, {Abell}, \& {Le
  Corre}}]{2014Icar..228..288S}
{Sanchez}, J.~A., {Reddy}, V., {Kelley}, M.~S., {et~al.} 2014, \icarus, 228,
  288

\bibitem[{{Scheinberg} {et~al.}(2015){Scheinberg}, {Fu}, {Elkins-Tanton}, \&
  {Weiss}}]{2015aste.book..533S}
{Scheinberg}, A., {Fu}, R.~R., {Elkins-Tanton}, L.~T., \& {Weiss}, B.~P. 2015,
  in Asteroids IV, ed. P.~{Michel}, F.~E. {DeMeo}, \& W.~F. {Bottke}, 533--552

\bibitem[{{Schulze} {et~al.}(1994){Schulze}, {Bischoff}, {Palme}, {Spettel},
  {Dreibus}, \& {Otto}}]{1994Metic..29..275S}
{Schulze}, H., {Bischoff}, A., {Palme}, H., {et~al.} 1994, Meteoritics, 29, 275

\bibitem[{{Scott} {et~al.}(2015){Scott}, {Keil}, {Goldstein}, {Asphaug},
  {Bottke}, \& {Moskovitz}}]{2015aste.book..573S}
{Scott}, E.~R.~D., {Keil}, K., {Goldstein}, J.~I., {et~al.} 2015, in Asteroids
  IV, ed. P.~{Michel}, F.~E. {DeMeo}, \& W.~F. {Bottke}, 573--595

\bibitem[{{Sunshine} {et~al.}(2007){Sunshine}, {Bus}, {Corrigan}, {McCoy}, \&
  {Burbine}}]{2007M&PS...42..155S}
{Sunshine}, J.~M., {Bus}, S.~J., {Corrigan}, C.~M., {McCoy}, T.~J., \&
  {Burbine}, T.~H. 2007, Meteoritics and Planetary Science, 42, 155

\bibitem[{{Sunshine} \& {Pieters}(1998)}]{1998JGR...10313675S}
{Sunshine}, J.~M. \& {Pieters}, C.~M. 1998, \jgr, 103, 13675

\bibitem[{{Sutherland} {et~al.}(2015){Sutherland}, {Emerson}, {Dalton},
  {Atad-Ettedgui}, {Beard}, {Bennett}, {Bezawada}, {Born}, {Caldwell}, {Clark},
  {Craig}, {Henry}, {Jeffers}, {Little}, {McPherson}, {Murray}, {Stewart},
  {Stobie}, {Terrett}, {Ward}, {Whalley}, \& {Woodhouse}}]{2015A&A...575A..25S}
{Sutherland}, W., {Emerson}, J., {Dalton}, G., {et~al.} 2015, \aap, 575, A25

\bibitem[{{Sykes} {et~al.}(2000){Sykes}, {Cutri}, {Fowler}, {Tholen},
  {Skrutskie}, {Price}, \& {Tedesco}}]{2000Icar..146..161S}
{Sykes}, M.~V., {Cutri}, R.~M., {Fowler}, J.~W., {et~al.} 2000, \icarus, 146,
  161

\bibitem[{{Tholen}(1984)}]{1984PhDT.........3T}
{Tholen}, D.~J. 1984, PhD thesis, University of Arizona, Tucson

\bibitem[{{{\v D}urech} {et~al.}(2005){{\v D}urech}, {Grav}, {Jedicke},
  {Denneau}, \& {Kaasalainen}}]{2005EM&P...97..179D}
{{\v D}urech}, J., {Grav}, T., {Jedicke}, R., {Denneau}, L., \& {Kaasalainen},
  M. 2005, Earth Moon and Planets, 97, 179

\bibitem[{{Veeder} {et~al.}(1982){Veeder}, {Matson}, \&
  {Kowal}}]{1982AJ.....87..834V}
{Veeder}, G.~J., {Matson}, D.~L., \& {Kowal}, C. 1982, \aj, 87, 834

\bibitem[{{Veeder} {et~al.}(1983){Veeder}, {Matson}, \&
  {Tedesco}}]{1983Icar...55..177V}
{Veeder}, G.~J., {Matson}, D.~L., \& {Tedesco}, E.~F. 1983, \icarus, 55, 177

\bibitem[{{Xu} {et~al.}(1995){Xu}, {Binzel}, {Burbine}, \&
  {Bus}}]{1995Icar..115....1X}
{Xu}, S., {Binzel}, R.~P., {Burbine}, T.~H., \& {Bus}, S.~J. 1995, \icarus,
  115, 1

\end{thebibliography}

\newpage
\clearpage
\onecolumn

\begin{appendix}
\section{Reference taxonomic system}
\label{Anexa1}

\begin{figure*}
\begin{center}
\includegraphics[width= \textwidth]{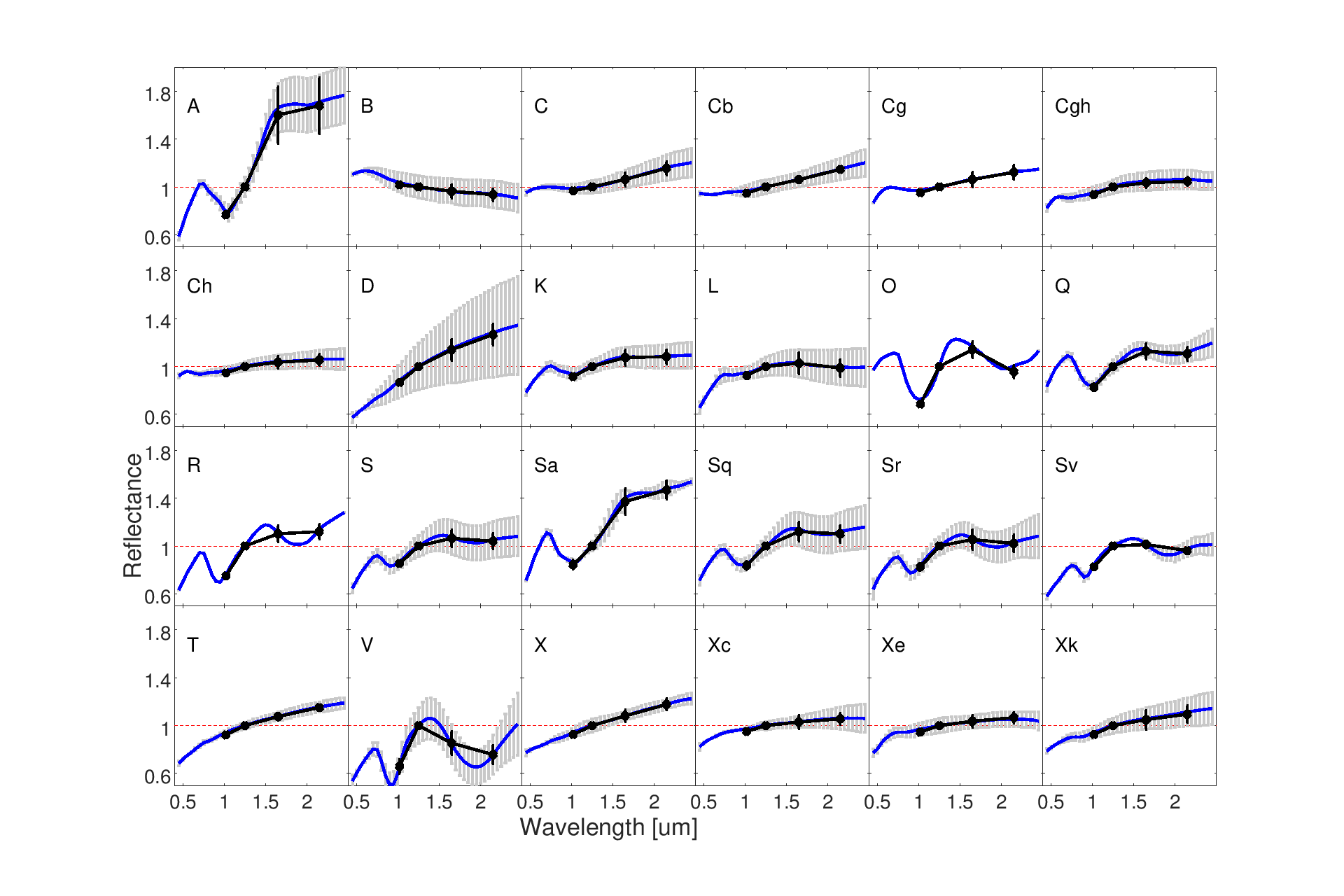}
\end{center}
\caption{ The template spectra (blue) and their standard deviations (grey) corresponding to taxonomic types defined by Bus-DeMeo taxonomy. Their equivalent reflectance for Y, J, H, and Ks VISTA filters are shown with black dots relative to their band centers.}
\label{ReflectanceClasses}
\end{figure*}

\begin{table*}[h!] 
\caption{  The average color values and their corresponding standard deviation of Bus-DeMeo taxonomic spectral types. The standard deviations were obtained from the computation of colors for the spectra used by \citet{2009Icar..202..160D} to define the taxonomy. A preliminary version of this table was shown by \citet{2016A&A...591A.115P}. The standard deviations, shown in this updated version, are limited to a minimum value of 0.02. We provide a standard deviation value (computed as median value of standard deviations of all classes) to the classes formed by a single object (Cg, O, R).}
\centering
\begin{tabular}{c c c c c c c c c c c c c c}\hline
Class& N &(Y-J)&$\sigma_{Y-J}$&(Y-H)&$\sigma_{Y-H}$&(Y-Ks)&$\sigma_{Y-Ks}$&(J-H)&$\sigma_{J-H}$&(J-Ks)&$\sigma_{J-Ks}$&(H-Ks)&$\sigma_{H-Ks}$\\ \hline
A&6&0.484&0.038&1.226&0.146&1.359&0.161&0.765&0.127&0.898&0.152&0.133&0.051      \\
B&4&0.177&0.020&0.368&0.050&0.417&0.072&0.214&0.035&0.263&0.058&0.049&0.023      \\
C&13&0.232&0.020&0.529&0.044&0.701&0.068&0.320&0.025&0.493&0.050&0.173&0.028      \\
Cb&3&0.254&0.020&0.549&0.011&0.715&0.030&0.318&0.020&0.484&0.021&0.166&0.020     \\
Cg&1&0.249&0.022&0.545&0.050&0.687&0.072&0.320&0.032&0.461&0.057&0.142&0.028     \\
Cgh&10&0.267&0.020&0.535&0.046&0.628&0.047&0.291&0.032&0.385&0.035&0.094&0.020    \\
Ch&18&0.256&0.022&0.525&0.044&0.626&0.061&0.292&0.026&0.393&0.048&0.101&0.026     \\
D&16&0.352&0.027&0.725&0.068&0.921&0.095&0.396&0.044&0.593&0.074&0.197&0.038      \\
K&16&0.292&0.026&0.601&0.065&0.690&0.079&0.332&0.043&0.421&0.059&0.089&0.024      \\
L&22&0.280&0.030&0.540&0.063&0.579&0.095&0.283&0.037&0.322&0.078&0.039&0.052      \\
O&1&0.606&0.022&0.980&0.050&0.870&0.072&0.397&0.032&0.287&0.057&-0.110&0.028     \\
Q&8&0.403&0.038&0.764&0.046&0.826&0.076&0.385&0.030&0.446&0.053&0.062&0.033      \\
R&1&0.507&0.022&0.844&0.050&0.943&0.072&0.360&0.032&0.459&0.057&0.099&0.028      \\
S&144&0.367&0.031&0.666&0.082&0.722&0.091&0.322&0.056&0.379&0.065&0.057&0.028      \\
Sa&2&0.381&0.049&0.956&0.053&1.111&0.020&0.598&0.020&0.753&0.057&0.156&0.061     \\
Sq&29&0.388&0.047&0.742&0.099&0.805&0.107&0.377&0.057&0.441&0.689&0.064&0.027     \\
Sr&22&0.404&0.039&0.693&0.080&0.740&0.096&0.311&0.048&0.359&0.075&0.048&0.039     \\
Sv&2&0.403&0.020&0.649&0.020&0.675&0.027&0.268&0.020&0.294&0.022&0.026&0.020     \\
T&4&0.284&0.020&0.594&0.020&0.750&0.020&0.333&0.020&0.489&0.020&0.157&0.020      \\
V&17&0.648&0.095&0.708&0.095&0.660&0.091&0.082&0.095&0.035&0.111&-0.047&0.050     \\
X&4&0.280&0.020&0.597&0.020&0.769&0.039&0.340&0.020&0.512&0.043&0.173&0.024      \\
Xc&3&0.251&0.020&0.513&0.025&0.624&0.057&0.285&0.020&0.396&0.047&0.111&0.034     \\
Xe&7&0.257&0.020&0.527&0.032&0.639&0.057&0.293&0.022&0.406&0.045&0.113&0.026     \\
Xk&18&0.280&0.020&0.564&0.050&0.692&0.086&0.307&0.034&0.435&0.071&0.129&0.039     \\
\hline
\end{tabular}
\label{TaxonColor}     
\end{table*}

\begin{figure*}
\begin{center}
\includegraphics[width=14cm]{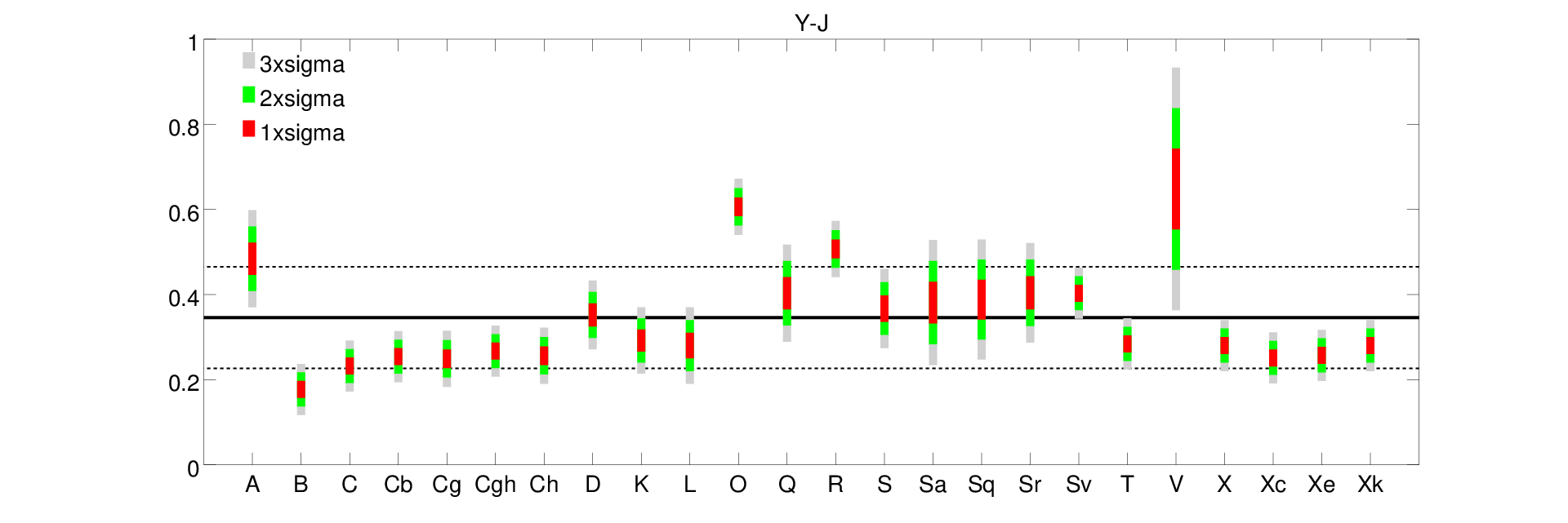}
\includegraphics[width=14cm]{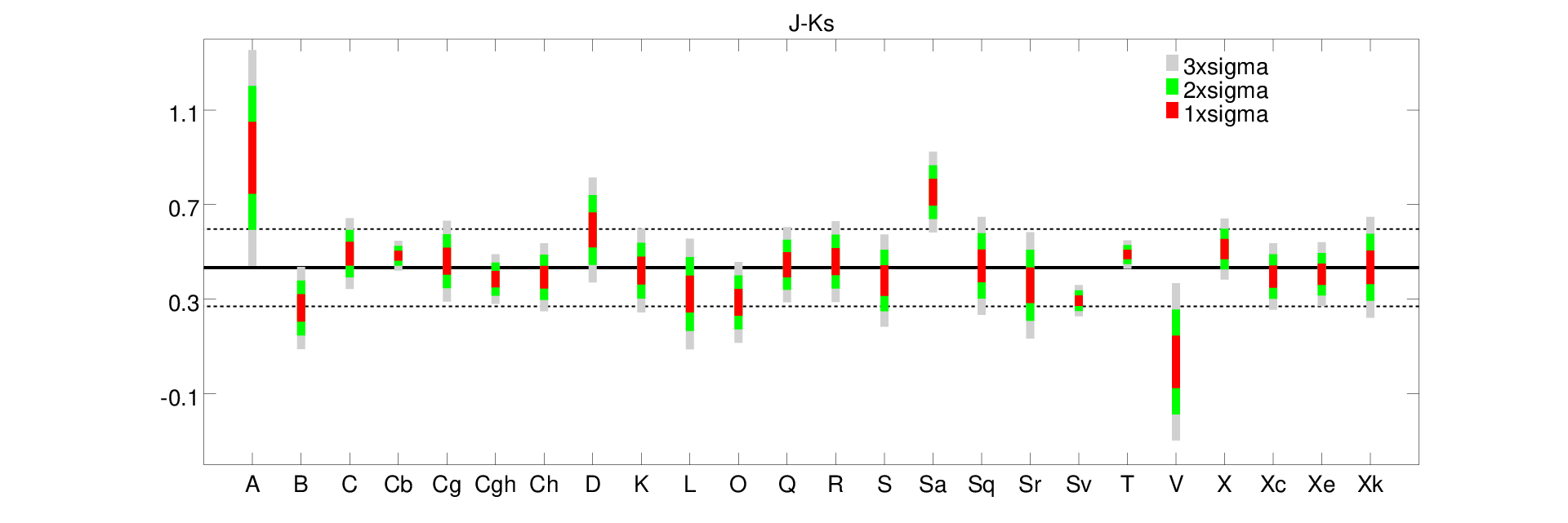}
\includegraphics[width=14cm]{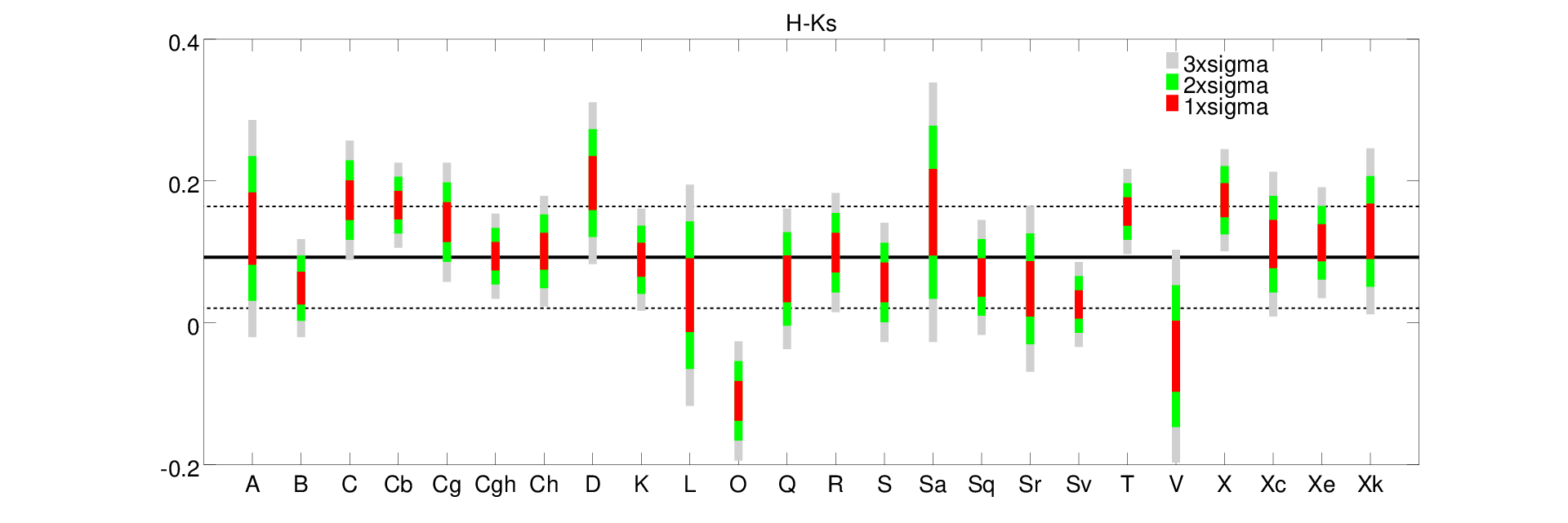}
\end{center}
\caption{The average colors and the standard deviations of Bus-DeMeo types computed from the 371 spectra from which the taxonomy was built. The red color shows the [-$\sigma$, +$\sigma$] interval, the green color shows [-$2\sigma$, +$2\sigma$], and the grey color shows [-$3\sigma$, +$3\sigma$]. The solid and dotted horizontal line represents the mean and the standard deviation of the entire set.}
\label{SingleColorDistrib}
\end{figure*}

\end{appendix}

\end{document}